\documentclass{svjour2}                    
\smartqed  
\usepackage[utf8]{inputenc}
\usepackage{ucs}
\usepackage{amsmath}
\usepackage{amsfonts}
\usepackage{amssymb}
\usepackage{graphicx}
\usepackage{subfigure}
\usepackage[french,english]{babel}
\usepackage{cite}
\journalname{J Stat Phys}
\begin{document}

\title{Random Convex Hulls and Extreme Value Statistics}



\author{Satya N. Majumdar \and Alain Comtet \and Julien Randon-Furling 
}

\institute{S. N. Majumdar \and A. Comtet \and J. Randon-Furling
\at
              LPTMS, Univ. Paris-Sud 11, UMR CNRS 8626,\\ 15 rue Georges 
Clemenceau,\\ 91405 Orsay Cedex, France \\ Universit\'e Pierre et Marie Curie-Paris
6,\\ 4 Place Jussieu,\\ 75252 Paris Cedex 05, France\\
              \email{majumdar@lptms.u-psud.fr}\\
              \email{comtet@lptms.u-psud.fr}
           \and
             \emph{Present address:} of J. Randon-Furling \at AG Rieger, Theoretische 
Physik,\\ Universität des Saarlandes, PF 151150,\\ 66041 Saarbrücken, Germany\\
             \email{randonfurling@lusi.uni-sb.de}
}

\date{Received: date / Accepted: date}

\maketitle

\begin{abstract}
In this paper we study the statistical properties of convex hulls
of $N$ random points in a plane chosen according to a given distribution.
The points may be chosen independently or they may be correlated.
After a non-exhaustive survey of the somewhat sporadic literature
and diverse methods used in the random convex hull problem, we 
present a unifying
approach, based on the notion of support function of a closed curve
and the associated Cauchy's formulae, that allows us to
compute exactly the mean perimeter and the mean area enclosed
by the convex polygon both in case of independent as well
as correlated points. 
Our method demonstrates a beautiful link
between the random convex hull problem and the subject
of extreme value statistics.
As an example of correlated points,
we study here in detail the case when the points represent the vertices
of $n$ independent random walks.
In the continuum time limit this reduces to 
$n$ independent planar Brownian trajectories for which 
we compute exactly, for all $n$, the 
mean perimeter and
the mean area of their global convex hull.
Our results
have relevant applications in ecology in estimating the home range 
of a herd of animals. Some of these results were announced
recently in a short communication [Phys. Rev. Lett. {\bf 103}, 140602 
(2009)].

\end{abstract}

\section{Introduction}

Convex sets are defined by the property that the 
line segment joining any two points of the set is itself fully 
contained in the set. In the physical world, convex shapes are encountered in 
many instances, from convex elements that ensure acoustic diffusion in concert halls \cite{Ros}, to crystallography, where the so-called Wulff construction leads, in the most general case, 
to a convex polyhedron whose facets correspond to crystal planes minimizing the surface energy \cite{Wu}. Also, recent work in neuroscience indicates that the human brain seems better able 
to distinguish between two distinct shapes when these are both convex \cite{Neuro}, which is particularly interesting since convexity properties are widely used in computer-aided image 
processing, in particular for 
pattern recognition \cite{AklTou}. Such applications would be limited if they were restrained to intrinsically convex shapes; but it is not the case, since it 
is possible to "approximate", in some sense, a non-convex object by a convex one: pick, among all convex sets that can enclose a given object, the smallest one in terms of volume. This is 
called its convex hull, and comparing convex hulls can be a viable mean of comparing the shapes of complex patterns such as proteins and docking sites \cite{Pro}. Convex hulls thus attract 
much interest, both for the algorithmic challenges set by their computation \cite{Grah,JAl,EdAl,PHAl,DevAL,KSAl,BSAl,WAl,Seidel, Tou2} and for their applications \cite{Pro, AklTou, CHMedic, 
CHIm, Worton}.
\medskip

Random convex hulls are the convex hulls of a set of
$N$ random points in a plane chosen according to some given distribution.
The points may be chosen independently each from an identical
distribution, e.g., from a uniform distribution over a disk.
Alternatively, the points may actually be correlated, e.g., 
they may represent the vertices of a planar random walk of
$N$ steps. For each realization of the set of points, one
can construct the associated convex hull. Evidently, the convex
hull will change from one realization of points to another.
Naturally, all geometric characteristics of the convex hull,
such as its perimeter, area, the number of vertices etc.
also become random variables, changing their values
from one realization of points to another.
The main problem that we are concerned here is to compute
the statistics of such random variables. For example, given
the distribution of the points, what is the distribution
of the perimeter or the area of the associated convex hull?    
It turns out that the computation of
even the first moment, e.g., the mean 
perimeter or the mean area of the convex hull
is a nontrivial problem.
\medskip

This question has aroused much interest among mathematicians over the past 50 
years or so, and has given rise
to a substantial body of literature some of which 
will be surveyed in Section 2. 
The methods used in this body of work turn out
to be diverse and sometimes specific to a given
distribution of points. It is therefore important
to find a unified approach that allows one
to compute the mean perimeter and area, both 
in the case of independent points as well as when
they are correlated such as in Brownian motion.
The main purpose of this paper is to present such
an approach. This approach is built on the
works of Tak\'{a}cs \cite{Ta}, Eddy \cite{Ed} and 
El Bachir \cite{El} and the main idea is to use
the statistical properties of a single object 
called the `support function' which allows us,
using the formulae known as Cauchy, 
Cauchy-Crofton or Cauchy-Barbier
formulae \cite{Cau, Cro, BaE, San, Val, AD},
to compute the mean perimeter and the mean area
of random convex hulls. This unified approach
allows us to reproduce the existing results obtained by
other diverse approaches, and in addition 
also provides new exact results, in particular for the 
mean perimeter and the mean area of $n$ independent planar Brownian motions
(both for open and closed paths),
a problem which has relevant applications in ecology in estimating
the home range of a herd of animals. The latter results
were recently anounced in a short Letter~\cite{RFSMAC}. 
\medskip

Our unified approach using Cauchy's formulae also establishes
an important link to the subject of extreme value statistics 
that deals with the study of the statistics of extremes
in samples of random variables. In the random convex hull 
problem when the sample points are independent and identically
distributed (i.i.d) in the plane, then the associated extreme value problem
via Cauchy's formulae is the classical example of extreme
value statistics (EVS) of i.i.d random variables that is
well studied, has found a lot of applications ranging
from climatology to oceanography and has a long history~\cite{EVSG1, 
EVSG2, Gne, EVSC}. For example, in the physics of disordered
systems the EVS of i.i.d variables plays an important role in the celebrated
random energy model~\cite{Derrida,BMM}.
In contrast, when the points are distributed in a correlated
fashion, as in the case where they represent the vertices
of a random walk, our approach requires the study
of the distribution of the maximum of a set of correlated variables, a subject
of much current interest in a wide range of problems (for a brief review
see~\cite{SMPK})
such as in fluctuating
interfaces~\cite{Burkhardt,Gyorgi1,Gyorgi2,Roy,ACSM1,SMAC,GSSM,Rambeau,Gyorgi2},
in logarithmically correlated Gaussian random energy models
for glass transition~\cite{CPL,FB1,Rosso1}, in the properties
of ground state energy
of directed polymers in a random media~\cite{Johansson,DS,DDSM1,PLD1,SMPK2,Sire1}
and 
the associated computer science problems on binary 
search trees~\cite{SMPK1,SMPK3,BKM} and the biological
sequence matching problems~\cite{Nechaev1},
in evolutionary dynamics and interacting particle 
systems~\cite{KJ,SMD,BenaM,SS,Mallick}, in loop-erased random walks~\cite{AgDh},
in queueing theory applications~\cite{KM}, in random jump processes
and their applications~\cite{CFFH,CMM,MZC}, in branching random
walks~\cite{Mckean,Bramson,BDer},
in condensation processes~\cite{EM},
in the statistics of records~\cite{Krug,MZ,GL,PLKW} and
excursions in nonequilibrium processes~\cite{GMS,Rosso,Sire},
in the density of near-extreme events~\cite{SSSM,SMR},
and also in various applications of the random matrix
theory~\cite{TW,DM1,DM2,BBP,VMB,Lak1,Lak2,SMCRF,MV1,Nad}.
Here, our approach establishes yet another application
of EVS, namely in the random planar convex hull problem. 

The paper is organised as follows. In Section 2, we provide a
non-exhaustive survey of the literature on random convex hulls.
In Section 3, we introduce the notion of the `support function'
and the associated Cauchy formulae for the perimeter and the area
of any closed convex curve. This section also
establishes the explicit link to extreme value statistics.
We show in Section 4 how to 
derive
the exact mean perimeter and the mean area of the convex hull of $N$
independent points using our approach. Section 5 is fully
devoted to the case when the points represent the vertices of
a Brownian motion, a case where the points are thus correlated.
Finally we conclude in Section 6 with some open questions.
Some of the details are relegated to the Appendices.

\section{A (non-exhaustive) review of results on 
random convex hulls}

In this section we briefly review a certain number of 
results on the the convex hull of 
randomly chosen points.
This review is of course far from exhaustive and we choose
only those results that are more relevant to this work.  
They are presented in a chronogical order and at the
end of the section we summarize in a table the results
that are particularly relevant to the present work.
\medskip

\begin{itemize}

\item In his book on random processes and Brownian motion (published in 1948) 
\cite{Lev}, P. Lévy mentions, in a few paragraphs and mainly
 heuristically the question of the convex hull of 
planar Brownian motion: "This contour [that of the 
convex hull of planar Brownian motion] consists, except for a null-measure set, in rectilinear parts."

\item More than ten years later, in 1959, J. Geffroy seems to be the 
first to publish results pertaining to the convex hull of a sample of 
random points drawn from a given distribution 
\cite{Ge1}, specifically $N$ points chosen in $\mathbb{R}^2$ according to a 
Gaussian normal distribution $f$. He shows that if one denotes\label{Geffr}
\begin{itemize}

\item by $\partial C _N$ the boundary of the convex hull of the sample,

\item by $\Sigma _N$ the ellipsoid given by the equation
\footnote{For instance, in 
the Gaussian case: $$f(x,y)=\exp\left[-\frac{x^2+y^2}{2}\right],$$ 
and the ellipsoid is simply the circle 
centered on the origin with radius $\sqrt{2\log N}$. We shall see further on how 
one can obtain directly the asymptotic behaviour of the perimeter of the convex hull in the case of $n$ 
points chosen at random in the plane according to a Gaussian normal distribution. 
It is given by:$$\langle L_N\rangle \sim 2\pi \sqrt{2\log N},$$ in complete 
agreement with Geffroy's 
result.} 
$f(x,y)=\frac{1}{N}$,

\item by $\Delta _N$ the distance between $\partial C _N$ and $\Sigma _N$,

\item and by $\Lambda _N$ the largest possible radius of an open disk whose 
interior lies inside the convex hull but contains none of the sample points,
\end{itemize}
then almost surely:
\begin{equation}
\Delta _N\text{ and }\Lambda _N  \mathop{\longrightarrow}_{N\rightarrow \infty} 0
\end{equation}
 In other words, the convex hull of the sample "tends" to the ellipsoid given by 
$f(x,y)=\frac{1}{N}$. Geffroy generalizes this result to $\mathbb{R}^d$, $d\geq 
1$, in 1961 \cite{Ge2}.

\item In 1961 F. Spitzer and H. Widom study the perimeter $L_N$ of the convex hull 
of a random walk represented by sums of random complex numbers $S_0=0,\cdots, 
S_k = Z_1 + Z_2 + \cdots Z_k  , 1\leq k\leq N $, where the
$Z_k$ are i.i.d. random variables. By combining an identity 
discovered by M.~K\'{a}c with a formula due to A.-L.~Cauchy 
(which  is used here for the first time in the context  of 
random convex hulls), they derive
an elegant formula for the expectation\cite{SW}:
\begin{equation}
\mathbb{E} (L_N) = 2 \sum_{k=1}^N \frac{\mathbb{E} (|S_k|)}{k} , \label{SW}
\end{equation}
The asymptotic behaviour of $\mathbb{E} (L_N)$ is studied in two different cases:
\begin{enumerate}

\item Writing $Z_k=X_k+iY_k$ and taking 
$\mathbb{E}(X_k) =\mathbb{E}(Y_k) =0$, $\mathbb{E}(X_k^2)=a^2$, 
$\mathbb{E}(Y_k^2)=b^2$, and $\mathbb{E}(X_kY_k)=\rho ab$, one has: 
\begin{equation}\mathbb{E}(L_N) \mathop{\sim}_{N\rightarrow \infty}  
4c\sqrt{N},\end{equation} where $c(a,b,\rho)$ does not depend on $N$.

\item Taking $Z_k=X_k+i$ with $\mathbb{E}(X_k)=\mu$ and 
$\mathbb{E}((X_k-\mu)^2)=\sigma^2$, one has: \begin{equation}
\mathbb{E}(L_N) \mathop{\sim}_{N\rightarrow \infty} 
2N\sqrt{1+\mu^2}+\frac{\sigma^2}{(1+\mu^2)^\frac{3}{2}}\log N,
\end{equation} which  expresses  the excess of $\mathbb{E} (L_N)$ over its 
smallest possible value
$2N\sqrt {1+\mu^2} $ .
\end{enumerate}

\item In the same year, and still for very general random walks 
viewed as a sum of $N$ vectors in the complex plane $S_0=0,\cdots, S_k = Z_1 + Z_2 
+ \cdots Z_k  , 1\leq k\leq N $, G. Baxter \cite{Bax} establishes three formulae 
involving respectively 
\begin{itemize}

\item the number $F_N$ of edges of the convex hull of the random walk,

\item  the number $K_N$ of steps $Z_k $ from the walk which belong to the boundary 
of the convex hull,

\item the perimeter $L_N$ of the of the convex hull .
\end{itemize}
In the latter case, Baxter's formula  coincides with Eq.~(\ref{SW}), but Baxter's 
derivation rests purely on combinatorial arguments and does not make use of 
Cauchy's formula. 
Instead, it relies on counting the number of permutations of the random walk's 
steps for which a given partial sum $ S_k $ belongs to 
the boundary of the convex hull.\\
His formulae are:
\begin{eqnarray}
\mathbb{E}(F_N)&=&2\sum_{m=1}^N \frac{1}{m}\sim 2\log N\\
\mathbb{E}(K_N)&=& 2\\
\mathbb{E} (L_N) &=& 2 \sum_{k=1}^N \frac{\mathbb{E} (|S_k|)}{k}.
\end{eqnarray}

\item In 1963, appears the first \cite{RS} of two seminal papers by  
A. Rényi and R. Sulanke dealing with the convex
hull of $N$ independent, 
identically distributed random points $P_i$ 
($i=1..N$) in dimension 2. Denoting by $F_N$ the number of edges of the convex 
hull, they consider the following cases:
\begin{enumerate}

\item the $P_i$'s are distributed uniformly within a convex, $r$-sided polygon~$K$:
\begin{equation}
\mathbb{E} (F_N) =  \frac{2}{3}r (\log N + \gamma) + T(K) + \text{o}(1)
\end{equation}
where $\gamma$ is the Euler constant and $T(K)$ is a constant depending on 
$K$ only and which is maximal for regular $r$-sided polygons,

\item the $P_i$'s are uniformly distributed within a convex set 
$K$ whose boundary is smooth:
\begin{equation}
\mathbb{E} (F_N) \mathop{\sim}_{N\rightarrow \infty} \alpha (K) N^\frac{3}{2},
\end{equation}
with $\alpha (K)$ a constant depending on $K$,

\item the $P_i$'s have a Gaussian normal distribution throughout the plane:
\begin{equation}
\mathbb{E} (F_N) \mathop{\sim}_{N\rightarrow \infty} 2\sqrt{2\pi \log N} 
\label{RS} 
\end{equation}
\end{enumerate}

\item The following year, 1964, the second of Rényi and Sulanke's
 papers \cite{RS2} extend these results, focusing on the asymptotic 
behaviour ${N\rightarrow \infty} $ of the perimeter $L_N$ and area $A_N$ of the 
convex hull of $N$ points $P_i$ drawn uniformly and independently within a convex 
set $K$ of perimeter $L$ and area $A$:
\begin{enumerate}

\item If $K$ has a smooth boundary:
\begin{eqnarray}
\mathbb{E} (L_N) &=& L - \mathcal{O}(N^{-\frac{2}{3}})\label{RSLnSm}\\
\mathbb{E} (A_N) &=& A - \mathcal{O}(N^{-\frac{2}{3}})\label{RSAnSm}
\end{eqnarray}

\item If $K$ is a square of side $a$:
\begin{eqnarray}
\mathbb{E} (L_N) &=& 4a - \mathcal{O}(N^{-\frac{1}{2}})\label{RSLnSq}\\
\mathbb{E} (A_N) &=& a^2 - \mathcal{O}(\frac{\log N}{N})
\end{eqnarray}
\end{enumerate}

\item In 1965, B. Efron \cite{Ef}, taking cue from Rényi and
Sulanke,
establish equivalent formulae in dimension 3, together with the average 
number of vertices (faces in dimension 3), 
the average perimeter and the average area of the convex hull of 
$N$ points drawn independently from a Gaussian normal distribution in dimension 2 
or 3, or from a uniform distribution 
inside a disk or sphere:
\begin{enumerate}

\item For instance, for $N>3$ points in the plane, drawn independently from a 
Gaussian normal distribution, writing 
$\phi (x)=(2\pi)^{-\frac{1}{2}} \exp (-\frac{1}{2}x^2)$ and $\Phi(x)= 
\int_{-\infty}^x \phi (y)\ dy$:
\begin{eqnarray}
\mathbb{E} (V_N) & = & 4 \sqrt{\pi} \binom{N}{2} \int_{-\infty}^\infty \Phi 
^{N-2}(p) \phi ^2(p)\ dp,\\
\mathbb{E} (L_N) & = & 4 \pi \binom{N}{2} \int_{-\infty}^\infty \Phi ^{N-2}(p) 
\phi ^2(p)\ dp,\label{EfLn}\\
\mathbb{E} (A_N) & = & 3 \pi \binom{N}{3} \int_{-\infty}^\infty \Phi ^{N-3}(p) 
\phi ^3(p)\ dp,\label{EfAn}
\end{eqnarray}
where $V_N$, $L_N$ and $A_N$ stand respectively for  the number of vertices,  
perimeter and area of the convex hull . 

\item In dimension 3, one has:
\begin{eqnarray}
\mathbb{E} (F_N) & = & 4 \sqrt{3\pi} \binom{N}{3} \int_{-\infty}^\infty \Phi 
^{N-3}(p) \phi ^3(p)\ dp,\\
\mathbb{E} (E_N)&=& \frac{3}{2} \mathbb{E} (F_N),\\
\mathbb{E} (V_N)&=& \frac{1}{2} \mathbb{E} (F_N) + 2,\\
\mathbb{E} (L_N) & = & 24 \sqrt{3\pi} \binom{N}{3} \int_{-\infty}^\infty \Phi 
^{N-3}(p) \phi ^3 (p)\ dp,\\
\mathbb{E} (A_N) & = & 12 \pi \binom{N}{3} \int_{-\infty}^\infty \Phi ^{N-3}(p) 
\phi ^3(p)\ dp,
\end{eqnarray}
$F_N$ and $E_N$ standing respectively for the number of faces and the number of 
edges of the convex hull ($L_N$ is thus the sum of the lengths of the edges, and 
$A_N$ the 
sum of the surface areas of the faces, $V_N$ denotes again the number of 
vertices).
\end{enumerate}
He also computes the average volume of the convex hull of $N$ vectors drawn 
independently from a Gaussian normal distribution 
(with zero average and unit variance) in a space of dimension 
$d < N$:
\begin{equation}
\mathbb{E} (\text{Vol}_N)= \frac{2\pi^{\frac{1}{2}d}}{\Gamma 
(\frac{1}{2}d)}\left(\frac{d+1}{d}\right)\binom{N}{d+1} \int_{-\infty}^\infty \Phi 
^{N-d-1}(p) \phi ^{d+1}(p)\ dp
\end{equation}
(for $N=d+1$, this expression needs to be multiplied by 2).

\item \begin{sloppypar} In 1965 still, H. Raynaud communicates in the 
\textit{Comptes Rendus de l'Académie des Sciences} \cite{Ray1}, his
 generalization to $\mathbb{R}^d$ of the formulae by 
Rényi-Sulanke and Efron pertaining to the number of vertices of
 the convex hull, either in the Gaussian normal case or in the uniform case. 
In the case of a Gaussian normal case, with 
zero mean and variance $a/2$, Raynaud computes the probability 
density of the convex hull of a sample and shows that, in the limit 
when the number of points $N$ in the sample becomes very 
large, this distribution converges to a uniform Poisson distribution 
on the sphere of radius $\sqrt{a\log N}$ centered at the origin.
\end{sloppypar}

\item
 In 1970, H. Carnal \cite{Ca} addresses the question of the convex hull of 
$N$ random points in the plane, drawn from a distribution which he assumes only to 
be circularly symmetric. 
He gives expressions for the asymptotic behaviour of the average number of 
edges, average perimeter and average area of the convex hull. In particular, 
he shows that the average number of 
edges, for certain distributions, goes to a constant (namely 4) 
when $N$ becomes very large.

\item H. Raynaud publishes in 1970 a second paper \cite{Ray2} on 
the convex hull of $N$ independent points (in both the Gaussian normal case 
throughout the space or the uniform case within 
a sphere) in $\mathbb{R}^d$. He gives detailed accounts of the results 
he announced earlier \cite{Ray1}. He also gives expressions for the 
asymptotic behaviour of the number of faces 
$F^{(d)}_N$ (or edges if $d=2$) of the convex hull, and shows in particular that 
in the standard Gaussian normal case:
\begin{equation}
\mathbb{E}(F^{(d)}_N) \mathop{\sim}_{N\rightarrow \infty} \frac{2^d}{\sqrt{d}} 
\left ( \pi \log N \right ) ^{\frac{1}{2}(d-1)}
\end{equation}
Note that for $d=2$, one recovers Rényi and Sulanke's formula 
Eq.(~\ref{RS}).
For $d=3$, one has $\mathbb{E}(F^{(3)}_N) \sim \frac{8}{\sqrt{3}}\pi \log N 
$

\item  Ten years later, in 1980, W. Eddy \cite{Ed} introduces the notion of 
support function into the field of random convex hulls. Considering, in the plane, 
$N$ points $P_i=(x_i,y_i)$  
with a Gaussian normal distribution, he associates to each a random process 
defined by: $$B_i(\theta)=x_i \cos \theta + y_i \sin \theta,$$ $\theta$ 
varying from $0$ to $2\pi$. 
Note that $B_i(\theta)$ is simply the projection of point 
$P_i$ on the line of direction $\theta$. He further defines the random process 
$$M(\theta)=\mathop{\sup}_i \{B_i(\theta)\},$$ whose law is  
shown to be given by that of the pointwise maximum of the 
$N$ independent, identically distributed random processes $B_i(\theta)$ (cf 
\cite{BR}). Eddy then shows that the point distribution of the 
stochastic process $M(\theta)$ is given by Gumbel's law, and he hints 
(without going further) to the fact that 
certain functionals of $M(\theta)$ give access to some geometrical 
properties of the convex hull of the sample: 
\begin{equation}
L_N =\int _0 ^{2\pi} M(\theta) d\theta \label{CA1}
\end{equation}
for the perimeter; and:
\begin{equation}
A_N =\frac{1}{2}\int _0 ^{2\pi} \left [ M^2(\theta) - (M'(\theta))^2 \right ] 
d\theta \label{CA2}
\end{equation}
for the area, where $M'(\theta)=\frac{dM}{d\theta}$. These functionals 
are what we refer to as Cauchy's formulae.

\item It is precisely the first of these formulae that 
L. Tak\'{a}cs\cite{Ta} suggests one should use to compute the 
expected perimeter length of the convex hull of planar Brownian motion, 
in his 1980 solution to a problem set 
by G. Letac in the \textit{American Mathematical Monthly} in 1978 \cite{Le1}. 
Denoting by  $L_t$ the perimeter of the convex hull of Brownian 
motion $ {B(\tau),0\leq \tau\leq t} $
, Tak\'{a}cs shows that:
\begin{equation}\mathbb{E} (L_t)  = \sqrt{8\pi t}.
\label{P1MB}\end{equation} The calculation is performed using the support 
function of the trajectory, in the same way as Eddy \cite{Ed} 
hinted at for independent points. Planar Brownian motion 
$B(\tau)$ is written as $(x(\tau),y(\tau))$, where $x$ and $y$ are 
standard one-dimensional Brownian motions. One then defines $$z_\tau 
(\theta)=x(\tau)\cos \theta + y(\tau) \sin \theta.$$ This stochastic 
process $z_\tau (\theta)$ being nothing else but the projection of the 
planar motion on the line with direction 
$\theta$, it is itself, for a fixed $\theta$, an instance of 
standard Brownian motion. Hence, the $M(\theta)$ that appears in 
Cauchy's formula (Eq.~(\ref{CA1})) and which is defined as: 
$$M(\theta)=\mathop{\max}_{0 \leq \tau \leq t} 
\{ z_\tau (\theta)\},$$ follows for a given $\theta$ the same law 
as the maximum of a standard one-dimensional Brownian motion. In particular, 
it is independent of $\theta$ and therefore one can write: 
$$ \mathbb{E}(L_t) = 2 \pi\ \mathbb{E}\left(M(0)\right),$$ 
thus using the isotropy of the distribution of planar Brownian motion. 
Knowledge of the right-hand part of this equation then yields the desired result.

\item In 1981, W. Eddy and J. Gale \cite{EG} extend further the work 
started by W. Eddy \cite{Ed} 
They point out the link between extreme-value statistics applied to $N$ 
1-dimensional random variables and the 
distribution of the convex hull of multidimensional random variables. 
They consider sample distributions with spherical symmetry and 
distinguish between three classes 
according to the shape of the tails: 
exponential, algebraic (power-law) or truncated (\textit{e.g.} 
distributed inside a sphere). Eddy and Gale then compute the asymptotic 
distribution of 
the associated stochastic process (the support function $M(\theta)$) 
when the number $N$ of points becomes very large. The three classes of initial 
sample distribution yield three types of 
distribution for the limit process, 
given by the Gumbel, Fréchet and Weibull laws,
 which are well known in the context of extreme-value statistics. 
Eddy and Gale also remark that the 
average number of vertices of the convex hull in the "Fréchet"
 case (that is, for initial sample distributions with power-law tails) 
tends to a constant (as proved by Carnal \cite{Ca}).

\item Following another route, N. Jewell and J. Romano establish the 
following year, in 1982, a correspondance between the random convex hull problem and 
a coverage problem, namely the covering  of the 
unit circle with arcs whose positions and lengths follow a bivariate law~\cite{JR}. 
Thus, for arcs of length $\pi$:\begin{center}
Prob (circle covered) $=$ Prob (conv. hull contains origin)
\end{center} and more generally, for arcs of lengths other than $\pi$: \begin{center}
Prob (circle covered) $=$ Prob (conv. hull contains a given disk)
\end{center}

\item In 1983, M. El Bachir, in his doctoral dissertation \cite{El}, 
studies the convex hull $C(t)$ of planar Brownian motion $B(t)$. 
In particular, he gives a proof of P. Lévy's assertion 
\cite{Lev} that almost surely $C(t)$ has a smooth boundary. 
El Bachir also shows that $C(t)$ is a Markov process on the set of compact convex domains containing the origin $O$. Denoting by 
$\partial C (t)$ the boundary of $C(t)$, he establishes:
\begin{enumerate}

\item $\mathrm{Prob}(B(t) \in \partial C (t)) = \mathrm{Prob}(O \in \partial C (t)) = 0$

\item $\{ t : B(t) \in \partial C (t)\}$ has a null Lebesgue measure .
\end{enumerate}
El Bachir then computes explicitly, from Cauchy's formulae, the expected perimeter length and surface area of the convex hull of planar Brownian motion. For the perimeter, he derives a 
general formula for motions with a drift $\mu$, of which the special case $\mu=0$ enables one to retrieve Tak\'{a}cs' $\sqrt{8\pi t}$. For the area, he obtains:
\begin{equation}
\mathbb{E}(A_t)=\frac{\pi t}{2}.\label{A1MB}
\end{equation}

\item Over the following decade, the study of the convex hull of a sample of 
independent, identically distributed random points has attracted much interest. 
C. Buchta \cite{B1} has obtained an exact 
formula giving the average area of the convex hull of 
$N$ points drawn uniformly inside a convex domain $K$, the existing formulae being 
so far mainly asymptotic. A few years later, F. 
Affentranger \cite{Aff}  has extended Buchta's result to higher dimensions, \textit{via} an induction relation. Many details and references can be found in the surveys of Buchta \cite{B2}, R. 
Schneider \cite{Sch}, W. Weil and J. Wieacker \cite{WW}.

\item Another active route is the one open by 
Eddy \cite{Ed} and Gale \cite{EG}, whose works have been 
extended  by H. Brozius and de Haan \cite{BH} to 
non-rotationally-invariant distributions. 
Brozius et al \cite{Bro}  also study the convergence in 
law to Poisson point processes exhibited by the distributions of 
quantities such as the number of vertices of the convex hull of 
independent, identically distributed random points. 
The works of Davis \textit{et al.}~\cite{Davis} and Aldous \textit{et 
al.}~\cite{Ald} 
also follow this type of approach.

\item Cranston \textit{et al.} \cite{CHM} 
resume the study of the convex hull $C(t)$ of planar Brownian motion and 
in particular of the continuity of its boundary $\partial C (t)$. They 
show that $\partial C (t)$ is almost surely $C^1$ and mention work 
by Shimura \cite{Shi1, Shi2} and K. Burdzy \cite{Bu} showing that for 
all $\alpha \in (\frac{\pi}{2},\pi)$, there exist 
random times $\tau$ such that $C(\tau)$ has corners of opening $\alpha$. 
They also mention Le Gall \cite{LeG} showing that the Hausdorff dimension of 
the set of times at which the Brownian 
motion visits a corner of $C(t)$ of opening $\alpha$ is almost 
surely equal to $1-\frac{\pi}{2\alpha}$. Finally, they also point 
to P. Lévy's paper \cite{Lev2} and S. N. Evas' \cite{Ev} 
for details on the growth rate of $C(t)$, as well as to 
K. Burdzy and J. San Martin's work \cite{KBSM} on the curvature of 
$C(t)$ near the bottom-most point of the Brownian trajectory.

\item In 1992, D. Khoshnevisan \cite{Kho} elaborates 
upon Cranston \textit{et al.}'s work by establishing an 
inequality that allows one to transpose, in some sense, the scaling properties 
of Brownian motion to its convex hull.

\item In 1993, two papers concerned with the convex hull of 
correlated random points, specifically the vertices of a random walk, 
are published. G. Letac \cite{Le2} points out  that Cauchy's 
formula enables one to write the  perimeter $L_N$ of the convex hull of any 
$N$-step random walk in terms of its support function 
$M_N(\theta)=\mathop{\max}_{0\leq i \leq N}\{x_i \cos 
\theta + y_i \sin \theta\}$:
\begin{equation}
\mathbb{E}(L_N)=\int_0^{2\pi}\ \mathbb{E}\left( M_N(\theta)\right)\ d\theta,
\end{equation}
which provides an alternative to Spitzer-Widom's or 
Baxter's methods to compute the perimeter of the convex hull of a random walk.\\

It is precisely to the Spitzer-Widom-Baxter's 
formula (Eq.~(\ref{SW})) that T. Snyder and J. Steele \cite{SnS} 
return, using again purely combinatorial arguments to obtain the 
following generalizations : \\
\begin{quotation}
Let $F_N$ be the number of edges of the convex hull of a 
$N$-step 
planar random walk and let $e_i$ be the length of the $i$-th edge. If 
$f$ is a real-valued function and if 
we set $G_N=\sum_{i=1}^{F_N}f(e_i)$, then:
\begin{equation}
\mathbb{E} (G_N) = 2 \sum_{k=1}^N \frac{\mathbb{E}\left(f\left(|S_k| 
\right)\right)}{k},\end{equation}  where   
  $ S_k = Z_1 + Z_2 + \cdots Z_k $ is the position of the walk after $k$ steps.\\
\begin{itemize}

\item Taking $f(x)=1$, one has $G_N=F_N$ (the number of edges of the convex 
hull) 
and one retrieves Baxter's result 
$$\mathbb{E}(F_N)=2\sum_{k=1}^N\frac{1}{k}\mathop{\sim}_{N\rightarrow 
\infty}2\log N.$$

\item Taking $f(x)=x$, one has $G_N=L_N$ and one retrieves Spitzer and Widom's 
result without using Cauchy's formula: $$\mathbb{E} (L_N) = 2 \sum_{k=1}^N 
\frac{\mathbb{E}\left(|S_k| \right)}{k}.$$

\item Taking $f(x)=x^2$, $G_N$ is the sum of the squared edges lengths denoted by 
$L_N^{(2)}$, one obtains: $$\mathbb{E}\left(L_N^{(2)}\right)= 2N(\sigma_X^2 
+\sigma_Y^2),$$ $\sigma_X^2+\sigma_Y^2$ being the variance of an 
individual step.\end{itemize}\end{quotation}
Snyder and Steele establish two other important results:
\begin{enumerate}

\item An upper bound for the variance 
$\mathbb{E}\left( L_N^2\right)$ (not to be mistaken for 
$\mathbb{E}\left(L_N^{(2)}\right)$) of the perimeter of the convex hull of any 
$N$-step planar 
random walk:
\begin{equation}
\mathbb{E}\left(L_N^2\right) \leq \frac{\pi ^2 }{2}N  (\sigma_X^2 +\sigma_Y^2)
\end{equation}

\item A large deviation inequality for the perimeter of the convex hull of an 
$N$-step random walk:
\begin{equation}
\mathrm{Prob}\left(|L_N-\mathbb{E}(L_N)|\geq t\right) \leq 2 e^{-\frac{t^2}{8\pi 
^2 N}}
\end{equation}
\end{enumerate}

\item In a paper published in 1996 \cite{Go}, A. Goldman 
introduces a new point of view on the convex hull of planar Brownian bridge. 
He derives a set of  new identities  relating the spectral empiral function of a 
homogeneous Poisson process to certain functionals of the convex hull. 
More precisely:\\
\begin{quotation} Let $D(R)$ be the open disk with radius $R$  and 
$D_i$ ($i=1..N_R$) the polygonal convex domains associated to a 
Poisson random measure and contained in $D(R)$. Let 
$$\phi_i(t)=\sum_{n=1}^\infty \exp(-t\lambda _{n,i})$$ be the 
spectral function of the domain $D_i$ (the $\lambda _{n,i}$'s being the 
eigenvalues of the Laplacian for the Dirichlet problem 
on $D_i$). Finally, set: $$\phi _R(t) = \frac{1}{N_R} \sum_{i=1}^{N_R}
\phi _i (t).$$ Then:
\begin{quotation}
\noindent $\phi_R(t)$ almost surely has a 
finite limit $\Phi(t)$ (called the empirical spectral function) 
when $R$ goes to infinity, and:
\begin{equation}
\Phi (t) = \frac{1}{4\pi ^2 t}\mathbb{E} \left( e^{-\sqrt{2t}L}\right)
\end{equation}
 where $L$ stands for the perimeter of the convex hull of 
the unit-time planar Brownian bridge (a Brownian motion 
conditionned to return to its origin at time $t=1$).
\end{quotation}
\end{quotation}
Goldman also computes the first  moment of $L$ using Cauchy's formula
\begin{equation}
\mathbb{E}(L)= \sqrt{\frac{\pi ^3}{2}}
\label{goldman-perimeter}
\end{equation}
To obtain the second moment,
\begin{equation}
\mathbb{E}(L^2)= \frac{\pi ^2}{3}\left (\pi \int_0^\pi \frac{\sin u}{u}\ du -1\right)
\end{equation}
 Goldman brings it down to computing $\mathbb{E}(M(\theta)M(0))$, 
the two-point correlation function of  the support function of the Brownian bridge,
\begin{equation}
\mathbb{E}(M(\theta)M(0))=\frac{\sin \theta}{2}\left[\frac{\theta(2\pi-\theta)}{6(\pi-\theta)}+\mathrm{cotan} \theta \right]
\end{equation}
Goldman obtains this last result from the probability that the Brownian bridge $\mathcal{B}_{0,1}$ lies entirely inside a wedge $\xi$ of opening angle $\beta$:
\begin{equation}
\mathrm{Prob}\left(\mathcal{B}_{0,1}\in \xi \right) = \frac{4\pi e^{-r^2}}{\beta}\sum_{k=1}^\infty \sin ^2 \left( \frac{k\pi \alpha}{\beta}\right) I_\nu(r^2),
\end{equation}
with $\nu=\frac{k\pi}{\beta}$, and, assuming that $O$ lies inside the 
wedge $\xi$, with $r$ the distance between $O$ and the apex $S$ of the wedge, and with $\alpha$ the angle between the line 
$O S$ and the closest edge of the wedge. ($I_\nu$ is the modified Bessel function of the first kind.)

\item In a later paper \cite{Go2}, Goldman exploits further the link between Poissonian mosaics and the convex hull of planar Brownian bridges. He first shows that one can replace Brownian 
bridges by simple Brownian motion. He then recalls Kendall's conjecture on Crofton's cell (in a Poisson mosaic, this is the domain $D_0$ that contains the origin): when the area $V_0$ of 
this cell is large, its "shape" would be "close" to that of a disk. Goldman shows in this paper a result supporting this claim (in terms of eigenvalues of the Laplacian for the Dirichlet 
problem) and, thanks to the links he has established, deduces that the convex hull of planar Brownian motion, when it is "small", has an "almost circular" shape. More precisely: if $C$ is 
the convex hull of a unit-time, planar Brownian motion $\mathcal{W}$, if $D(r)$ is the disk centered at the origin with radius $r \in (0,\infty)$ and if $M=\sup \{||W(s)||,0\leq s \leq 
1\}$, then, for all $\epsilon \in (0,1)$:
\begin{equation}
\limsup_{a\rightarrow 0} \mathrm{Prob}[D((1-\epsilon)a) \subset C \subset D(a)|M=a]=1
\end{equation}

\item In parallel, refined studies of the asymptotic distributions and 
limit laws of the number of vertices, perimeter or area of the 
convex hull of independent points drawn uniformly inside a 
convex domain $K$ continue, in particular with the work of 
P. Groeneboom \cite{Gro} on the number of vertices 
(augmented by S. Finch and I. Hueter's result \cite{FiH}), Hsing \cite{Hs} on 
the area when $K$ is a disk, Cabo and Groeneboom \cite{CaGro} 
for the area too but when $K$ is polygonal, Bräker and Hsing \cite{BHs}
 for the joint law of the perimeter and area, and the more recent 
works of Vu \cite{Vu2}, Calka and Schreiber \cite{Cal}, 
Reitzner \cite{Re1,Re2,Re3} and B\'{a}r\'{a}ny \textit{et al.}\cite{Bar,BaRe,Vu1}.

\item In 2009, P. Biane and G. Letac \cite{BiLe} return to the convex 
hull of planar Brownian motion, focusing on the global convex hull of 
several copies of the same trajectory (the copies 
obtained via rotations). They compute the expected perimeter length of this global 
convex hull for various settings. 
\end{itemize}

Thus we see that random convex hulls have aroused much interest over the past 
50 years or so. The main results relevant to our present study are 
those giving explicit expressions (exact or 
asymptotic) for the average perimeter and area of the convex hull of a 
random sample in the plane. We have attempted to group the 
corresponding references in the following table:
\begin{center}
\begin{tabular}{|p{2.0cm}|p{0.8cm}||p{3.4cm}|p{3.4cm}|}
\hline
\multicolumn{2}{|l||}{\textsc{Existing results}} & 
\textbf{Perimeter} (average) & \textbf{Area}  (average) \\
\hline
\hline
\multicolumn{2}{|l||}{} &\multicolumn{2}{c|}{}\\
\multicolumn{2}{|l||}{\textbf{Independent}}& 
\multicolumn{2}{c|}{\small \textit{Rényi and Sulanke \cite{RS2}} 
Eq.~(\ref{RSLnSm}) and (\ref{RSLnSq}));}\\ \multicolumn{2}{|l||}{ }& 
\multicolumn{2}{c|}{\small\textit{Efron \cite{Ef}} (Eq.~(\ref{EfLn}) et 
(\ref{EfAn})); \textit{Carnal \cite{Ca}};}\\
\multicolumn{2}{|l||}{\textbf{Points}} &\multicolumn{2}{c|}{ \small \textit{Buchta \cite{B1}; Affentranger \cite{Aff}}}\\
\multicolumn{2}{|l||}{} &\multicolumn{2}{c|}{}\\
\hline
&&&\\
\begin{center}\textbf{Random walk}\end{center}&\begin{center} open 
paths \end{center} & \small \textit{Spitzer et Widom \cite{SW}} 
(Eq.~(\ref{SW}));\newline \textit{Baxter \cite{Bax}}; \newline 
\textit{Snyder et Steele \cite{SnS}}; \newline \textit{Letac \cite{Le2}}&\begin{center} ? \end{center} \\&&&\\ \cline{2-4}
\begin{center}\textbf{(1 walker)}\end{center}& \begin{center}closed 
paths\end{center} &\begin{center} ? \end{center} &\begin{center} ? 
\end{center}\\ \hline
\begin{center}\textbf{Brownian motion}\end{center}& \begin{center}open paths 
\end{center}& \small \begin{center}  \textit{Takács \cite{Ta}} 
Eq.~(\ref{P1MB}))\end{center}& \begin{center}  \small 
\textit{El Bachir} \cite{El} (Eq.~(\ref{A1MB}))\end{center}\\ \cline{2-4}
\begin{center}\textbf{(1 motion)}\end{center} & \begin{center} closed paths 
\end{center}& \begin{center} \small\textit{Goldmann \cite{Go}} 
(Eq.~(\ref{goldman-perimeter})) \end{center} & \begin{center} ? 
\end{center}\\
\hline
\end{tabular}
\end{center}
\bigskip

Finally, we have developed a general method recently~\cite{RFSMAC}
that enabled us not only to 
fill in 
the empty cells in this table but also to treat a generalization which is 
particularly relevant physically, namely the geometric properties of the 
global convex hull of $n>1$ independent planar Brownian paths, each of
the same duration $T$. To the best 
of our 
knowledge, this topic had never been addressed before. Among the works 
mentioned 
above, those dealing with correlated points always consider a single random 
walk or a single Brownian motion, except for \cite{BiLe} where several copies of 
the 
same Brownian path are considered.

Yet, the convex hull of several random paths appears quite naturally, both on 
the theoretical side and also in the context of ecology, as we shall see later 
(\S~\ref{ApplEco}). Furthermore, the simplest case, that of $n$ independent 
Brownian motions, already exhibits interesting features since the geometry of 
the 
convex hull depends in a non trivial manner on $n$~\cite{RFSMAC}. We showed 
that even 
though
the Brownian walkers are independent, the global convex hull of the
union of their trajectories depend on the multiplicity $n$
of the walkers in a nontrivial way. In the large $n$ limit,
the convex hull tends to a circle with a radius $\sim \sqrt{\ln 
n}$~\cite{RFSMAC}
which turns out to be identical to that of the set of
distinct sites visited by $n$ independent random walkers on a 2-dimensional
lattice~\cite{Larralde,LarrNature,Yuste}. 
This general method will be developed in detail in the following sections.

\section{Support function and Cauchy formulae:\\ a general approach to random 
convex hulls}

In this section we discuss the notion of the `support function'.
Intuitively speaking, the support function in a certain direction
$\theta$ of a given two dimensional object is the maximum spatial
extent of the object along that direction. We will see that the
knowledge of this function for all angles $\theta$ can be
fruitfully used to obtain the perimeter and the area of any
closed convex curve (in particular for a convex polygon)
by virtue of Cauchy's formulae.

\subsection{Support function of a closed convex curve and Cauchy's formulae}

Let $C$ denote any closed and smooth convex curve in a plane. For example $C$ may 
represent 
a circle or an ellipse. The curve $C$ may be represented by the coordinates
of the points on it $\{(X(s),Y(s))\}$ parametrized by a continuous $s$.
Associated with $C$, one can construct a support function in a natural
geometric way. Consider any arbitrary direction from
the origin $O$ specified by the angle $\theta$ with respect to the $x$ axis.
Bring a straight line
from infinity perpendicularly along direction $\theta$ and
stop when it touches a point on the curve $C$. The support function
$M(\theta)$, associated with curve $C$, 
denotes the 
Euclidean (signed) distance of this perpendicular
line from the origin when it stops, measuring the maximal
extension of the curve $C$ along the direction $\theta$.
\begin{equation}
M(\theta)= \max _{s\in C} \left\{X(s) \cos \theta +Y(s)
\sin\theta\right\}.  
\label{M0}
\end{equation}

The knowledge of $M(\theta)$ enables one to compute the perimeter of $C$ and also 
the area enclosed by $C$ via Cauchy's formulae 
\begin{eqnarray} L 
&=&\int_{0}^{2\pi}d\theta M(\theta) \label{C1}\\ A 
&=&\frac{1}{2}\int_{0}^{2\pi}d\theta \left( M^2 (\theta) - 
\left(M'(\theta)\right)^2\right).
\label{C2} 
\end{eqnarray} 
These formulae are 
straightforward to establish for polygonal curves, and taking the continuous limit 
in the polygonal approximation yields the result for smooth curves (a
non-rigorous, but quick, `proof' is provided in appendix A).

\begin{figure}[h]
\begin{center}
\subfigure[]{\includegraphics[width=4.1cm,height=4cm]{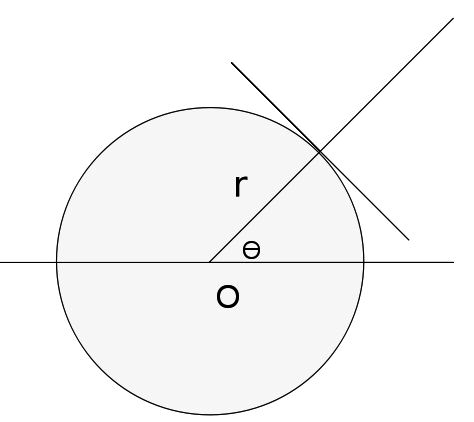}\label{Circle}}
\subfigure[]{\includegraphics[width=4cm,height=4cm]{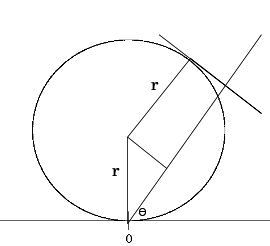}\label{CircleO}}
\caption{Simple examples for Cauchy formulae: circle centered on the origin, 
and circle "resting" on the origin}
\end{center}
\end{figure}

In the elementary example of a circle centered on the origin with 
radius $r$ (Fig.~\ref{Circle}), 
$M(\theta)$ is constant and equal to $r$ for all $\theta$. Its derivative is zero and 
Cauchy formulae give the standard results. 
The second example is slightly less 
trivial (Fig.~\ref{CircleO}) as $M(\theta)$ is not constant but 
equal to $r(1+\sin\theta)$. One of course 
recovers again the usual results:
\begin{eqnarray*}
L&=&\int_{0}^{2\pi}d\theta\ r(1+\sin\theta) =2\pi r\\
A&=&\frac{1}{2}\int_{0}^{2\pi}d\theta\ r^2 
[(1+\sin \theta)^2- \cos ^2 \theta] = \pi r^2.
\end{eqnarray*}

It is interesting to note that Cauchy's original motivations
for deriving the formulae \eqref{C1} and \eqref{C2} actually
came from a somewhat different context. He was interested
in developing a method to compute the roots of certain algebraic equations 
as a convergent series and to compute an upper bound of the
error made in stopping the series after a finite number of terms.
It was in this connection that he proved a number of theorems
concerning the length of the perimeter and the area enclosed
by a closed convex curve in a plane. Anticipating the
usefulness of his formulae in a variety of contexts and particularly in
geometrical applications, he collected them in a self-contained "Memoire"
published by the "Academie des Sciences" in 1850. It is worth
pointing out that Cauchy's formulae are of purely geometric origin
without any probabilistic content. The idea of using these formulae
in probabilistic context was first used by Crofton~\cite{Cro} whose work
can be considered as one of the starting points of the subject
of {\em integral geometry}, developed by Blaschke and his school
during the years 1935-1939.

\subsection{Support function of the convex hull of a discrete set of points in 
plane}

Let $I=\{(x_i,y_i), i=1,2,\ldots, N \}$ denote a set of $N$ points
in a plane with coordinates $(x_i,y_i)$. 
Let $C$ denote the
convex hull of $I$, i.e., the minimal convex polygon enclosing this set.
This convex hull $C$ is a closed, smooth convex curve and hence we can
apply Cauchy's formulae in Eqs. (\ref{C1}) and (\ref{C2}) to compute
its perimeter and area. However, to apply these formulae we first need
to know the support function $M(\theta)$ associated with the convex hull
$C$, as given by Eq. (\ref{M0}). This requires a knowledge of the coordinates 
$(X(s),Y(s))$ of a point, parametrized by $s$, on the convex polygon $C$.
A crucial point is that one can compute this support function
associated with the convex hull $C$ of $I$ just from the knowledge
of the coordinates $(x_k,y_k)$ of the set $I$ itself and without
requiring first to compute the coordinates $(X(s), Y(s))$ of the points
{\it on} the convex hull $C$. Indeed the support function
$M(\theta)$ associated with $C$ can be written as
\begin{equation}
M(\theta)= \max_{s\in C}\left\{X(s) \cos \theta +Y(s)
\sin\theta\right\}= \max _{i \in I} \left\{x_i \cos \theta +y_i 
\sin\theta\right\}.
\label{M}
\end{equation}
This simply follows from the fact that $M(\theta)$, the maximal extent of 
the convex polygon $C$ along $\theta$, is also the maximum of the
projections of all points of the set $I$ along that direction $\theta$.
Thus, the knowledge of the coordinates $(x_k,y_k)$ of any set $I$
is enough to determine the support function $M(\theta)$ of the
convex hull $C$ associated with $I$ by Eq. (\ref{M}).
\begin{figure}[h]
\begin{center}
\includegraphics[height=4cm, width=6cm]{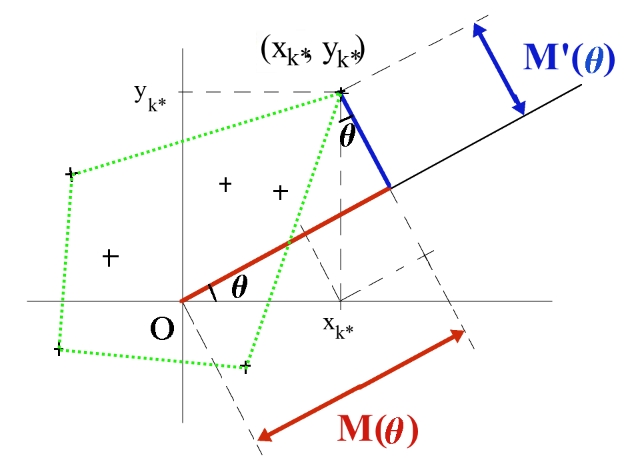}
\end{center}
\caption{Support function $M(\theta)$ and its derivative $M'(\theta)$ 
of the convex hull (dotted green lines) assocaited with a set
of $7$ points.}
\label{SupFct2}
\end{figure}

We also note that, 
by definition of $M(\theta)$ in Eq. (\ref{M}), for any fixed $\theta$ 
there will be a point $(x_{k^*}, y_{k^*})$ in the set such that:
\begin{equation}
M(\theta)= x_{k^*} \cos \theta +y_{k^*} \sin\theta 
\label{M1}
\end{equation}

Taking derivative of (\ref{M1}) with respect to $\theta$ gives
\begin{equation}
M'(\theta)= -x_{k^*} \sin \theta +y_{k^*} \cos\theta \label{M'1}
\end{equation}
In other words, $M'(\theta)$ is the distance 
between the point of the set giving the maximal 
projection $M(\theta)$ and the straight line with 
direction $\theta$, as illustrated in 
Figure~\ref{SupFct2}.

\subsection{A Simple illustration of the support function $M(\theta)$ of
a triangle}

\begin{figure}[!h]
\begin{center}
\includegraphics[height=4.5cm, width=4.90cm]{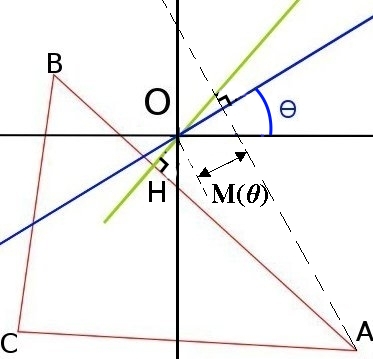}
\end{center}
\caption{Triangle $ABC$ (in red), with the line of direction $\theta$ (in blue) and the line through $O$ perpendicular to $AB$}\label{Tr}
\end{figure}

\begin{figure}[h]
\begin{center}
\subfigure[]{\includegraphics[height=5cm, width=7.5cm]{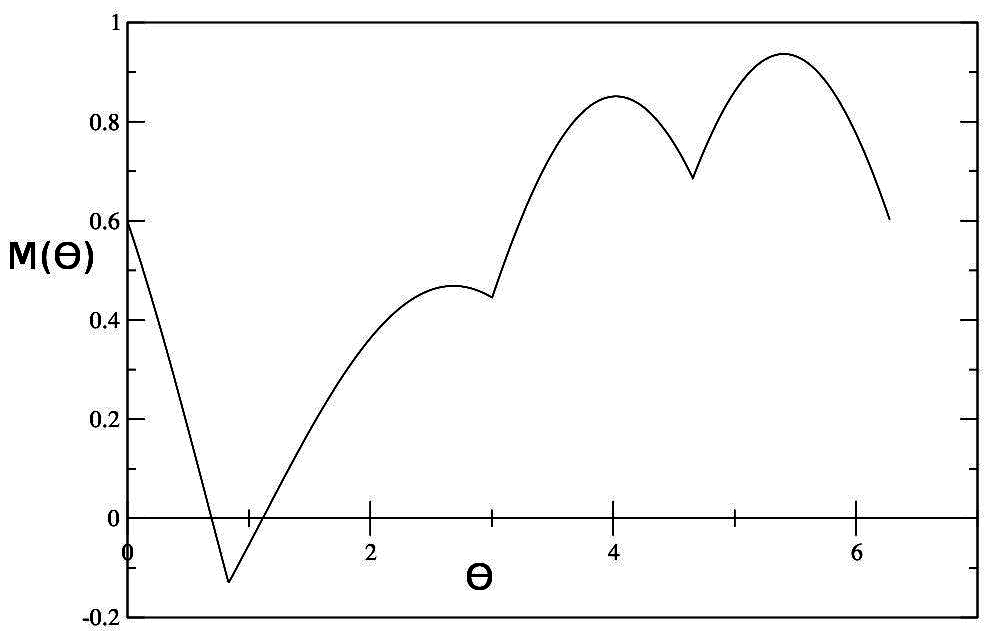}\label{MTr}}
\subfigure[]{\includegraphics[height=5cm, width=7.5cm]{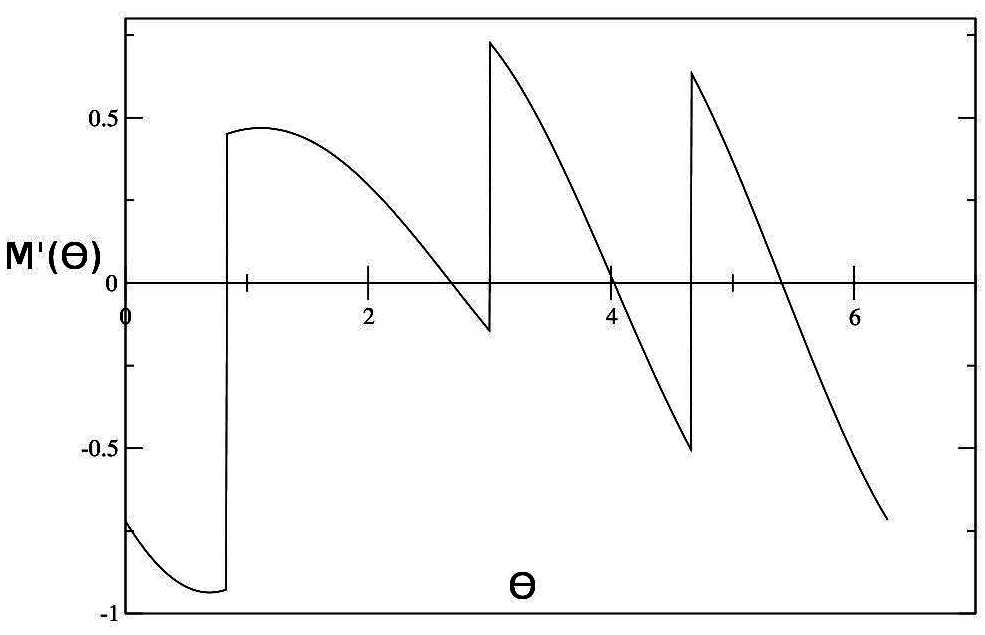}\label{dMTr}}
\end{center}
\caption{(a) Support function $M(\theta)$  of triangle $ABC$ and (b) its derivative $M'(\theta)$ }
\end{figure}

To get familar with the support function $M(\theta)$ associated with
a convex hull, let 
us consider a simple 
example of three points in a plane. The associated convex hull
is evidently a triangle  
(Fig.~\ref{Tr}) whose support 
function $M(\theta)$ and derivative $M'(\theta)$ are drawn 
in  Figures~\ref{MTr} and \ref{dMTr}.

$M(\theta)$ is of course $2\pi$-periodic. A notable feature of its graph is the presence of angular points, corresponding to discontinuities of the derivative of $M(\theta)$. So, 
$M(\theta)$ appears piecewise smooth, with a derivative exhibiting a 
finite number of jump discontinuities. This finite number is, in 
the special case considered here, equal to three, the 
number of points in the set whose support function is $M(\theta)$. 
This is not by chance and the coincidence can be understood by returning to 
equations~(\ref{M1}) and (\ref{M'1}): within a 
given range of $\theta$, one of the vertices of the triangle $ABC$ will be giving the maximal projection on direction $\theta$ and will thus determine the value of $M(\theta)$, say:
\begin{equation}
M(\theta)=x_{A} \cos \theta + y_{A} \sin \theta ,\label{MA}
\end{equation}
Then, within the same range of $\theta$:
\begin{equation}
M'(\theta)=-x_A \sin \theta + y_A \cos \theta. \label{DMA}
\end{equation}
We can specify the range of angles $\theta$ on which equations (\ref{MA} and \ref{DMA}) are valid. Let us indeed note that by the definition of $M(\theta)$, when $\theta$ 
corresponds to the perpendicular through the origin $O$ to the line segment $AB$,  $A$ and $B$ have the same projection on direction $\theta$ (Fig.~\ref{Tr}). In the $\theta^-$-limit, that 
is for angles approaching $\theta$ from below, the value of $M$ will be given by the projection of $A$, and the value of $|M'(\theta)|$ by the length of the line segment $AH$ ($H$ being 
the foot of the perpendicular to $AB$ through $O$). In the $\theta^+$-limit, for angles slightly larger than $\theta$, the value of  $M(\theta)$ will still be the common projection of $A$ 
and $B$ but $|M'(\theta)|$ will be given by the length $BH$. Whence, as $\theta$ passes on the perpendicular to $AB$ through $O$, the support function $M$ will be continuous, while its 
derivative will have a jump discontinuity.

Indeed, looking at figure~\ref{Tr}, and starting from $\theta=0$, we see that point $A$ gives the maximal projection on 
direction $\theta$, and the length of this projection decreases as $\theta$ increases, until the direction given by $\theta$ coincides with line $(OH)$, which is the perpendicular to 
$[AB]$ through the origin. At this point, as we have just noticed, $M(\theta)$ has an angular point: $B$ will then give the maximal projection, whose value will increase until $\theta$ 
corresponds to line $(OB)$ where $M(\theta)$ attains a local maximum before decreasing until $\theta$ coincides with the perpendicular to $[BC]$ through $O$, where $M(\theta)$ has a second 
angular point; and so on.\footnote{In the specific example chosen here, all 3 vertices of the triangle are "visible" through a local maximum of $M(\theta)$. However, this is not always the 
case. This is easily seen by considering a configuration in which point $H$ (the foot of the perpendicular to $(AB)$ through $O$), while being by definition on the line $(AB)$, is not on 
the line segment $[AB]$: for example, if $H$ is beyond $A$, $A$ will go "unnoticed".  Yet, the coincidence of direction $\theta$ with line $(OH)$ will always result in a discontinuity of 
$M'(\theta)$ (although without change in sign), which corresponds to an angular point for $M(\theta)$. Thus the angular points of $M(\theta)$ count the number of sides (and, in dimension 2, 
of vertices) of the convex hull. As for the local maxima of $M(\theta)$, they only count the number of vertices $E$ of the convex hull that are such that the maximal projection on line 
$(OE)$ is given by $E$ itself ---~one could call such vertices "extremal" or "self-extremal" vertices.}
\medskip

\subsection{Cauchy formulae applied to a random sample}

Let us now examine how the Cauchy formulae can be applied to 
determine the mean perimeter and the mean area
of a convex hull associated with a set of $N$ points with coordinates
$(x_i,y_i)$ in
a plane chosen from some underlying probability distribution. The 
points may be independent or correlated.  
\medskip

For each $i$ and fixed $\theta$, let us define
\begin{eqnarray}
z_i(\theta)&=&x_i \cos \theta + y_i \sin \theta \label{zeq}\\
h_i(\theta)&=&-x_i \sin \theta + y_i \cos \theta. \label{heq}
\end{eqnarray}
$z_i$ is simply the projection of the $i$-th point in the sample on direction 
$\theta$ and $h_i$ its projection on the direction perpendicular to $\theta$. By 
definition (Eq.~(\ref{M})): \begin{equation}
M(\theta) = \mathop{\max}_{i } \{ z_i (\theta)\}\equiv z_{k^*}(\theta)
\end{equation}
for a certain index $k^*$.\\
One then has:
\begin{equation}
M'(\theta)=h_{k^*}(\theta)
\end{equation}
\medskip

When the points $(x_i,y_i)$ are random variables, so is the index $k^*$
and subsequently both $M(\theta)$ and $M'(\theta)$ are also
random variables. Taking averages in Cauchy's formulae (\eqref{C1})
and (\eqref{C2}) we get the mean perimeter and the mean area 
\begin{eqnarray}
\langle L\rangle &=&\int_{0}^{2\pi}d\theta \ \langle M(\theta)\rangle \label{AC1} \\
\langle A\rangle &=&\frac{1}{2}\int_{0}^{2\pi}d\theta \ 
\left( \langle M^2 (\theta)\rangle - 
\langle \left(M'(\theta)\right)^2\rangle\right)\label{AC2}
\end{eqnarray}
where $\langle \cdot \rangle$ 
indicates an average over all realisations of the points, and we assume that this 
operation commutes with the integration over $\theta$.
\medskip

In the most general setting, let us also define
\begin{itemize}

\item $\mu_\theta$ be the probability density function of the maximum of the 
$z_i(\theta)$, i.e., of the random varibale $z_{k^*}(\theta)$

\item $\rho_{\theta}$ be the probability density function of the index $k^*$
for which $z_i(\theta)$ becomes the maximum 

\item and $\sigma_{i,\theta}$ be the probability density function of the 
random variable $h_i(\theta)$ for a fixed $i$ and $\theta$,
\end{itemize}
then:
\begin{eqnarray}
\langle M(\theta)\rangle &= &\int_{-\infty}^{\infty}\ z\ \mu _\theta (z)\ dz\label{MM1}\\
\langle M^2(\theta)\rangle &=& \int_{-\infty}^{\infty}\ z^2\ \mu _\theta (z)\ dz\label{MM2}\\
\langle \left(M'(\theta)\right)^2\rangle &= &\int_I  \int_{-\infty}^{\infty}\ h^2\ \rho_\theta(k)\ \sigma _{k,\theta} (h)\ dk\ dh \\
&=&\int_I\ \rho_\theta(k)\ \langle h_k^2(\theta)\rangle\ dk \label{MM3}
\end{eqnarray}
\medskip

With this formulation, it appears explicitly that random convex hulls are 
directly linked with extreme-value statistics, the study of extremes in samples 
of random variables. Indeed, when $I$ is finite and the $N$ points labelled by 
$i\in I$ are chosen independently and are identically distributed, for instance in 
$\mathbb{R}^2$, then $\mu_{\theta}$ is the distribution of the maximum of $N$ 
real-valued i.i.d random variables (namely the $z_i(\theta)$'s) ---~a classical example 
of EVS~\cite{EVSG1, EVSG2, Gne, EVSC}. One can then use directly the results
of the standard EVS of i.i.d random variables. 
On the other hand, when the points are correlated, we need
to study the distribution of the maximum of a set
of correlated random variables--a subject of much current interest
as mentioned in the introduction. Here we need to go beyond 
i.i.d variables and take into acount the strong correlations
between the random variables that changes the distribution
of their maximum in a nontrivial way.

\medskip

If the sample points are the vertices of an $N$-step 2-dimensional random walk, then the $z_i(\theta)$'s can be seen, for a fixed $\theta$ as the vertices of an $N$-step 1-dimensional 
random walk, and $\mu_{\theta}$ is the distribution of the maximum of such a walk. Note that in this case, $\rho_\theta$ is the distribution of the step at which the 1-dimensional random 
walk $z_i(\theta)$ attains its maximum \cite{DanSky, Odly, CMY, Feller}.
\medskip

One can also consider cases when $I$ is not a discrete, finite set: \textit{e.g.} 
the random set might be the trajectory $\mathcal{B}(\tau)=(x(\tau),y(\tau))$ of a 
planar Brownian motion at 
times $\tau \in I=[0,T]$. In such a case, both $z_\tau(\theta)$ and $h_\tau(\theta)$ are instances of 1-dimensional Brownian motion, and $\mu_\theta$ is the distribution of the maximum of 
1-dimensional Brownian motion in $[0,T]$, $\rho_{\theta}$ is the distribution of 
the time at which 1-dimensional Brownian motion attains its maximum in $[0,T]$ 
(given by Lévy's arcsine law 
\cite{Lev3}), and $\sigma _{\tau,\theta}$ the propagator of 1-dimensional Brownian motion between $0$ and $\tau$ (\textit{i.e.} the distribution of the position of a linear Brownian motion 
after a time $\tau$).
\medskip

In the following section, we use this approach to compute the mean perimeter
and the mean area of the convex hull of a set of $N$ indepedently chosen
points in a plane. In Section 5, we will examine how the same approach 
can be adapted to compute the mean perimeter and the mean area 
of the convex hull of $n$ independent planar Brownian paths each of
the same duration $T$.

\section{Independent Points}
\subsection{General case} 

Let us consider here a sample of $N$ points drawn independently from a bivariate distribution:
$${\rm Prob} \left(x_i\in[x,x+dx], y_i\in[y,y+dy]\right)=p(x,y)\ dx\ dy,$$

Following the route explained in the previous section (Eqs.~(\ref{zeq}), 
(\ref{heq})), we let:
$$z_i(\theta)=x_i \cos \theta + y_i \sin \theta,$$and:
$$h_i(\theta)=-x_i \sin \theta + y_i \cos \theta
$$

\subsection{Isotropic cases}
Let $(x_1, y_1), (x_2, y_2), \cdots, (x_N, y_N)$ be 
$N$ points in the plane, each drawn independently 
from a bivariate distribution $p(x,y)$ that is invariant
under rotation, i.e., $p(x,y)=G(\sqrt{x^2+y^2})$.
In such an isotropic case, the distribution of the 
support function $M(\theta)$  does not depend on $\theta$ and 
it is thus sufficient to set $\theta=0$ and hence
$M_N\equiv M(0)$. The random 
variables $z_i(0)$ and $h_i(0)$ are just, respectively, 
the abscissa $x_i$ and ordinate $y_i$ of the 
points. Combining (\ref{AC1}) and (\ref{MM1}), 
we can then write the average perimeter of the 
convex hull
\begin{equation}
\langle L_N \rangle = 2\pi \langle \mathop{\max}_{i}\{x_i\}\rangle \equiv 2\pi \langle M_N\rangle \label{ACM}
\end{equation}
It is useful to first define the cumulative distribution
\begin{equation}
F_N(M) = \text{Prob} [M_N \leq M].
\end{equation}
For independent variables it follows that
\begin{eqnarray}
F_N(M)=\left[\int_{-\infty}^M\ p_X(x)\ dx\right]^N,
\end{eqnarray}
where $p_X(x)=\int_{-\infty}^\infty\ p(x,y)\ dy$ is the 
marginal of the first variable $X$.
Thus, in a general isotropic case
\begin{eqnarray}
\langle M_N\rangle &=& \int_{-\infty}^\infty\ M\ F_N'(M)\ dM\nonumber\\
\langle L_N\rangle &=&2\pi\ N\int_{-\infty}^\infty\ M\ p_X(M) \left[\int_{-\infty}^M\ p_X(x)\ dx\right]^{N-1}\ dM\nonumber\\
&=&2\pi\ N\int_{-\infty}^\infty\ M\ p_X(M)\ F_{N-1}(M)\ dM\label{LIs}
\end{eqnarray}

For the average area in the isotropic case, we can write it as (Eqs.~(\ref{AC2}), 
(\ref{MM2}), (\ref{MM3})):
\begin{equation}
\langle A_N\rangle = \pi \langle M_N^2\rangle -\pi \langle y_{k^*}^2\rangle, \label{AIs}
\end{equation}
where $y_{k^*}$ is the ordinate of the point $(x_{k^*},y_{k^*})$ with the largest abscissa, \textit{i.e.} satisfying:
$$x_{k^*} =\mathop{\max}_{i}\{x_i\}=M_N.$$
We can easily express the second moment of $M_N$ that appears in (\ref{AIs}):
\begin{eqnarray}
\langle M_N^2\rangle &=& \int_{-\infty}^\infty\ M^2 F_N'(M)\ dM\\
&=&N\int_{-\infty}^\infty\ M^2\ p_X(M)\ F_{N-1}(M)\ dM.\label{Mn2}
\end{eqnarray}
To compute the second term in (\ref{AIs}), that is, the second moment of the ordinate of the point with largest abscissa, we first compute the probability density function $\hat{p}$ of this 
point, which is defined by:
\begin{equation}
\text{Prob}\ \left\{(x_{k^*},y_{k^*})\in \left[(x,y),(x+dx,y+dy)\right]\right\}= \hat{p}(x,y)\ dx\ dy
\end{equation} 
(Note that $F_N(M)=\int\ \hat{p}(M,y)\ dy$.)
\medskip

\noindent It is not difficult to see that $\hat{p}(x_{k^*},y_{k^*})$ can be expressed as the probability density that one of the $N$ points has coordinates $(x_{k^*},y_{k^*})$ and the $N-1$ 
other points have abscissas less than $x^*$:
\begin{equation}
\hat{p}(x_{k^*},y_{k^*}) = N\ p(x_{k^*},y_{k^*})\ \left[\int_{-\infty}^{x_{k^*}}\ p_X(x)\ dx\right]^{N-1}
\end{equation}
Then:
\begin{equation}
\langle y_{k^*}^2\rangle = N\ \iint_{-\infty}^\infty\ y_{k^*}^2\ p(x_{k^*},y_{k^*})\ F_{N-1}(x_{k^*})\ dx_{k^*}\ dy_{k^*}\label{M'2}
\end{equation}
It  now suffices to insert (\ref{Mn2}) and (\ref{M'2}) in (\ref{AIs}) to obtain a general expression for the average area of the convex hull of $N$ points drawn independently from an 
isotropic bivariate distribution $p$ with marginal $p_X$:
\begin{multline}
\langle A_N\rangle = N\ \pi\ \int_{-\infty}^\infty\ u^2\ p_X(u)\ F_{N-1}(u)\ du\\
-N\ \pi\ \iint_{-\infty}^\infty\ v^2\ p(u,v)\ F_{N-1}(u)\ du\ dv \label{AIs2}
\end{multline}

The equations (\ref{LIs}) and (\ref{AIs2}) are the main results
of this subsection. They provide the exact mean perimeter and the mean
area of the convex hull of $N$ independent points in a plane each
drawn from an arbitrary isotropic distribution. As an example,
let us consider the case of a Gaussian distribution where
the general expressions can be further simplified. 
Let
\begin{equation}
p(x,y)= \frac{1}{2\pi} e^{-\frac{1}{2}(x^2+y^2)}.
\end{equation}
We then have:
\begin{equation}
p_X(x)=\frac{1}{\sqrt{2\pi}}\exp \left(-\frac{x^2}{2}\right)\equiv \phi(x)
\end{equation}
and:
\begin{equation}
\int_{-\infty}^x\ p_X(x')\ dx' = \int_{-\infty}^x\ \phi(x')\ dx' \equiv \Phi(x)
\end{equation}

Inserting these into equations (\ref{LIs}) and (\ref{AIs}), and performing suitable integrations by parts, we obtain:
\begin{eqnarray}
\langle L_N\rangle &=& 4\ \pi\ \binom{N}{2}\ \int_{-\infty}^\infty\ \Phi ^{N-2}(x)\ \phi ^2 (x)\ dx\\
\langle A_N\rangle &=& 3\ \pi\ \binom{N}{3}\ \int_{-\infty}^\infty\ \Phi 
^{N-3}(x)\ \phi ^3 (x)\ dx
\end{eqnarray}
which coincide with the expressions derived by Efron \cite{Ef} using a 
rather different method.

\subsection{Asymptotic behaviour of the average perimeter and area}

To derive how the mean perimeter and the mean area behave for
large $N$, we need to investigate the asymptotic large $N$
behavior of the two exact expressions in Eqs. (\ref{LIs}) and (\ref{AIs}).
For the mean perimeter, since it is exactly identical to the
maximum $M_N$ (upto a factor $2\pi$) of $N$ independent
variables each distributed via the marginal $p_X(x)$, we can
use the standard analysis used in EVS, which is summarized below.
For the mean area, on the other hand, we need to go further.
We will give a specific example of this asymptotic analysis of the
mean area later.

\medskip

\begin{quotation}
\begin{center}
\textbf{Summary of standard extreme-value statistics}
\end{center}
\bigskip

\begin{sloppypar}\noindent Let $z_1, z_2,\dots, z_N$ be independent, identically distributed  random variables with probability density function $p(z)$, and let $M_N=\max 
_{\kappa =1..N} \left\{z_\kappa \right\}$ be their maximum. Then
$$ F_N(M)=\text{Prob} \left(M_N \leq M\right)=\left[\int_{-\infty}^M\ p(z)\ dz \right]^N. $$
In the limit when $N$ becomes very large, the cumulative 
distribution function $F_N(M)$ exhibits one of the three 
following behaviours, according to the shape of the "tails" of the parent 
distribution $p(z)$:\end{sloppypar}
\medskip

\begin{enumerate}

\item When the random variable $z$ has unbounded support and its distribution
$p(z)$ has a faster than power law tail as $z\to \infty$. We will loosely
refer to this as ``Exponential tails". Then, 
"Exponential tails lead to a Gumbel-type law"  $$p(z) \mathop{\sim}\limits_{z \rightarrow \infty}  A\ e^{-z^{\alpha}}\ \rightarrow\quad F_N(M) \mathop{\sim}\limits_{N \rightarrow 
\infty} e^{-e^{-(M^\alpha-\log N)}} $$

\item "Power-law tails lead to a Fréchet-type law" $$p(z)  \mathop{\sim}\limits_{z \rightarrow \infty} A\ z^{-(\alpha + 1)}\ \rightarrow\quad F_N(M) \mathop{\sim}\limits_{N \rightarrow 
\infty} e^{-\frac{A}{\alpha}\ N\ M^{-\alpha}}$$

\item "Truncated tails (\textit{i.e.} finite range $a$) lead to a Weibull-type law (with parameter $a$)" $$p(z) \mathop{\sim}\limits_{z \rightarrow a} A\ (a-z)^{\alpha -1}\ 
\rightarrow\quad F_N(M) \mathop{\sim}\limits_{N \rightarrow \infty} e^{-\frac{A}{\alpha}\ N\ (a- M)^{\alpha}}$$
\end{enumerate}
\end{quotation}
In all three cases, the typical value of the maximum $M_N$ increases
as $N$ increases\footnote{As 
$\log N$ in the first case, as a power of $N$ in the second, and 
nearing as an inverse power of $N$ the radius $a$ of the interval in the third case.}:
the larger the number of points, the 
further the maximum is pushed.
\bigskip

We will use these results for the general asymptotic behavior of the mean 
perimeter. However, before providing a summary of the asymptotic behavior
for a general isotropic distribution, it is perhaps useful to 
consider two special cases in detail, one for the mean perimeter and one for 
the mean area, that will illustrate how one can carry out this
asymptotic analysis. For the mean perimeter, we choose the
parent distribution from the Weibull-type case and for the mean area
we choose the Fr\'echet-type distribution. These choices
are somewhat arbitrary, one could have equally chosen any other case
for illustration.
\medskip

\noindent \textbf{Example 1: average perimeter in the "Weibull-type" case}

\begin{quotation}
\noindent Let us consider $N$ points drawn independently inside a circle of radius $a$ from a distribution with Weibull-type tails:
\begin{equation}
p(x,y)\mathop{\sim}_{\sqrt{x^2+y^2}\rightarrow a} A\ 
(a-\sqrt{x^2+y^2})^{\gamma-1}.
\end{equation}
\medskip

\noindent Letting as before $F_N$ denote the cumulative distribution function of the maximum $M_N$ of the $x$-coordinates, an integration by parts yields:
\begin{eqnarray}
\langle M_N \rangle &=& \int_{-a}^a\ x\ F_N'(x)\ dx\nonumber\\
&=& a-\int_{-a}^a\ F_N(x)\ dx\label{Mnn}
\end{eqnarray}
\medskip

\noindent We focus on the second term of (\ref{Mnn}) and write: $$I_N=\int_{-a}^a\ F_N(x)\ dx.$$
Then:
\begin{eqnarray}
I_N &=& \int_{-a}^a \left[ 1-\int _x ^a\ p_X(x')\ dx'\right]^N\ dx\\
&=& \int_{-a}^a \exp \left[N\ \log \left( 1-\int _x ^a\ p_X(x')\ dx'\right)\right]\ dx,
\end{eqnarray}
where as before we write $p_X(x)=\int\ p(x,y)\ dy$.
\medskip

\noindent We now pick $0<\epsilon \ll 1$ such that $$\text{for}\quad (a-\epsilon)<x<a,$$ $$p(x,y)\simeq A\ (a-\sqrt{x^2+y^2})^{\gamma-1}.$$ The idea being that when $N$ becomes large, some 
sample points will come closer and closer to the boundary (the circle of radius $a$) and consequently one can focus on the tails of the distribution. We therefore write:
\begin{equation}
I_N=I_N^{(1)}+I_N^{(2)}
\end{equation}
with:
\begin{equation}
I_N^{(1)}=\int_{-a}^{a-\epsilon} \exp \left[N\ \log \left( 1-\int _x ^a\ p_X(x')\ dx'\right)\right]\ dx
\end{equation}
and
\begin{equation}
I_N^{(2)}=\int_{a-\epsilon}^{a} \exp \left[N\ \log \left( 1-\int _x ^a\ p_X(x')\ dx'\right)\right]\ dx\label{I2}
\end{equation}
\medskip

\noindent It is possible to show that $I_N^{(1)}$ decreases exponentially with $N$ and is, as expected on heuristic grounds, subleading compared to $I_N^{(2)}$ which, as we are going to 
see, decreases as an inverse power in $N$.
\medskip

\noindent The sample distribution $p(x,y)$ is rotationally
invariant and bounded ($x^2+y^2\le a^2$). Hence the marginal
\begin{eqnarray}
p_X(x)&=&\int_{-\sqrt{a^2-x^2}}^{\sqrt{a^2-x^2}}\ p(x,y)\ dy\\
&=&2\ \int_{0}^{\sqrt{a^2-x^2}}\ p(x,y)\ dy
\end{eqnarray}
Now for $x\lesssim a$, we have: $p(x,y)\sim A\ (a-\sqrt{x^2+y^2})^{\gamma-1}$. Consequently, setting $y=xu$ and considering that $(a-\epsilon)<x<a$:
\begin{eqnarray}
p_X(x)&\sim & 2\ \int_{0}^{\sqrt{a^2-x^2}}\ A\ (a-\sqrt{x^2+y^2})^{\gamma-1}\ dy\\
&\sim & 2Ax\ \int_{0}^{\sqrt{\frac{a^2}{x^2}-1}}\ (a-x\ \sqrt{1+u^2})^{\gamma-1}\ du\\
&\sim & 2Aa\ \int_{0}^{\sqrt{\frac{2(a-x)}{a}}}\ (a-x)^{\gamma-1}\ du\\
&\sim & 2\ A\ \sqrt{2a}\ (a-x)^{\gamma-\frac{1}{2}}
\end{eqnarray}
We now proceed from equation~(\ref{I2}):
\begin{eqnarray}
I_N^{(2)}&=&\int_{a-\epsilon}^{a} \exp \left[N\ \log \left( 1-\int _x ^a\ p_X(x')\ dx'\right)\right]\ dx\nonumber\\
&\sim & \int_{a-\epsilon}^{a} \exp \left[N\ \log \left( 1-\int _x ^a\ 2\ A\ \sqrt{2a}\ (a-x')^{\gamma-\frac{1}{2}}\ dx'\right)\right]\ dx\nonumber\\
&\sim & \int_{a-\epsilon}^{a} \exp \left[N\ \log \left( 1-\frac{4\ A\ \sqrt{2a}}{2\ \gamma + 1}\ (a-x)^{\gamma+\frac{1}{2}}\right)\right]\ dx\nonumber\\
&\sim & \int_{a-\epsilon}^{a} \exp -\left[\frac{4\ A\ N\ \sqrt{2a}}{2\ \gamma + 1}\ (a-x)^{\gamma+\frac{1}{2}}\right]\ dx
\end{eqnarray}
To progress further, we perform the following change of variable:
\begin{equation}
u=\frac{4\ A\ N\ \sqrt{2a}}{2\ \gamma + 1}\ (a-x)^{\gamma+\frac{1}{2}}
\end{equation}
In the large $N$ limit in which we are working, this change of variable leads to:
\begin{eqnarray}
I_N^{(2)}&\sim & \frac{2}{\left[4AN\sqrt{2a}\right]^{\frac{2}{1+2\gamma}}}\int_0^\infty\ e^{-u}\ \left[(2\gamma +1)\ u\right]^{\frac{1-2\gamma}{1+2\gamma}}\ du\nonumber\\
&\sim & \frac{2 \left(2\gamma +1\right)^{\frac{1-2\gamma}{1+2\gamma}}\ \Gamma \left( \frac{2}{2\gamma +1}\right)}{\left[4AN\sqrt{2a}\right]^{\frac{2}{1+2\gamma}}}\label{In2R}
\end{eqnarray}

\noindent The combination of (\ref{In2R}) with (\ref{Mnn}) and (\ref{ACM}) yields the final result:
\begin{equation}
\langle L_N \rangle \mathop{\sim}_{n\rightarrow \infty}2\pi a - \frac{4\ \pi \left(2\gamma +1\right)^{\frac{1-2\gamma}{1+2\gamma}}\ \Gamma \left( \frac{2}{2\gamma 
+1}\right)}{\left[4AN\sqrt{2a}\right]^{\frac{2}{1+2\gamma}}}
\end{equation}
\medskip

\noindent To illustrate this asymptotic result for a concrete example, 
consider $N$ points drawn independently and uniformly from a unit disk
\begin{equation}
p(x,y)=\frac{1}{\pi}\ \Theta (1-x^2-y^2),
\end{equation}
where $\Theta$ is the Heaviside step function.

\noindent In terms of our notations, this corresponds to:
\begin{eqnarray}
a&=&1,\\
A&=&\frac{1}{\pi},\\
\gamma &=& 1.
\end{eqnarray}
We find:
\begin{equation}
\langle L_N \rangle \mathop{\sim}_{N\rightarrow \infty}2\pi\ \left(1-\frac{\Gamma \left(\frac{2}{3}\right)\pi^{\frac{2}{3}}}{12^\frac{1}{3}\ N^\frac{2}{3}}\right),
\end{equation}
in complete agreement with Rényi and Sulanke's result Eq.~(\ref{RSLnSm})) 
\cite{RS2}.
Note that for large $N$, the mean perimeter of the convex hull approaches $2\pi$,
i.e., the convex hull approaches the bounding circle of radius unity of the 
disk.
But it approaches very slowly, the correction term decreases for large $N$
only as a power law $\sim N^{-2/3}$. Actually, for this example of
uniform distribution over a unit disk, one can also obtain
simple and explicit expressions for the mean perimeter and the mean area
starting from our general expressions in Eqs. (\ref{LIs}) and (\ref{AIs}).
Skipping details, we get
\smallskip

\noindent Perimeter:
\begin{equation}
\langle L_N \rangle = 2\pi\ \left[1 - \int_{-1}^1\ F_N (M)\ dM \right]
\end{equation}
\smallskip

\noindent Area:
\begin{equation}
\langle A_N \rangle = \pi\ \left[1 - \frac{8}{3}\int_{-1}^1\ M\ F_N (M)\ dM \right]
\label{ANWeib}
\end{equation}
\smallskip

\noindent where:
\begin{equation}
F_N(M)=\frac{1}{\pi}\ \left[ \arcsin (M) + M\ \sqrt{1-M^2}\right]^N
\label{CumWeib}
\end{equation}
\medskip

One can also easily work out the asymptotic behavior of the mean area
in this example using Eqs. (\ref{ANWeib}) and (\ref{CumWeib}) and we get
\begin{equation}
\langle A_N \rangle \mathop{\sim}_{N\rightarrow \infty}\pi\ 
\left(1-2\frac{\Gamma \left(\frac{2}{3}\right)2^\frac{7}{3}
\pi^{\frac{2}{3}}}{3^\frac{4}{3}\ N^\frac{2}{3}}\right)
\end{equation}
which, once again, agrees with Rényi and Sulanke's result 
(Eq.~(\ref{RSAnSm})) \cite{RS2}. Note also that the mean area 
of the convex hull approaches, for large $N$, to the area
of the unit disk. 
Notice also that the exponent of $N$, 
which governs the speed of convergence is the same 
for the area as for the perimeter ---~only the prefactor of the power of $N$ 
changes\footnote{Rényi and Sulanke \cite{RS2} have shown that
 this is in fact true for every smooth-bounded support, and, 
moreover, with the same universal exponent: $N^{-\frac{2}{3}}$.}.

At this point it is also worth recalling the results of Hilhorst \textit{et al.} 
\cite{Hilhorst} regarding Sylvester's problem \footnote{If $N$ points are drawn from a uniform distribution in the unit disk, what 
is the probability $p_N$ that they be the vertices of a convex polygon ---~in other words that they be the vertices of their own convex hull?}. When $N$ becomes large, the convex hull of the 
$N$ points (conditioned to have all the $N$ points to be its vertices) lies in an 
annulus of width $\sim N^{-\frac{4}{5}}$ smaller than the $N^{-\frac{2}{3}}$ 
found in our case. This can be understood qualitatively by noticing that requiring the $N$ points to be on the convex hull will tend to increase the size of the hull and therefore push it closer 
to the boundary of the disk.
\end{quotation} 
\bigskip

\noindent \textbf{Example 2: average area in the "Fréchet-type" case}

\begin{quotation}
Consider $N$ points drawn independently from an isotropic distribution 
with Fréchet-type tails:
\begin{equation}
p(x,y)\mathop{\sim}_{\sqrt{x^2+y^2}\rightarrow \infty} 
\frac{A}{(x^2+y^2)^\frac{\gamma +2}{2}}.
\end{equation}
\medskip

\noindent Recalling equation (\ref{AIs2}), we start by its first term. 
Letting as before $F_N$ denote the cumulative distribution function of the 
maximum $M_N$ of the $x$-coordinates, we 
have:
\begin{equation}
\langle M_N^2 \rangle = \int_{-\infty}^\infty\ x^2\ F_N'(x)\ dx\equiv I_N\label{MnA}.
\end{equation}
With the same notation as previously:
\begin{equation}
F_N(x)=\left[ 1-\int _x ^\infty\ p_X(x')\ dx'\right]^N
\end{equation}
where as before we write $p_X(x)=\int\ p(x,y)\ dy$.
\medskip

\noindent We pick $K \gg 1$ such that for $$x\geq K,$$ $$p(x,y)\simeq \frac{A'}{(x^2+y^2)^\frac{\gamma +2}{2}}.$$ The idea being that when $N$ becomes large, sample points will disseminate 
further and further in the plane, and consequently one can focus on the tails of the distribution. We therefore write:
\begin{equation}
I_N=I_N^{(1)}+I_N^{(2)}
\end{equation}
with:
\begin{equation}
I_N^{(1)}=\int_{-\infty}^{K}\ x^2\ F_N'(x) \ dx
\end{equation}
and
\begin{equation}
I_N^{(2)}=\int_{K}^{\infty}\ x^2\ F_N'(x) \ dx\label{Ia2}
\end{equation}
\medskip

\noindent It is easy to show that $I_N^{(1)}$, as before,
is subleading compared to $I_N^{(2)}$.
\medskip

\noindent The sample distribution $p(x,y)$ is rotationally invariant and so:
\begin{eqnarray}
p_X(x)&=&\int_{-\infty}^{\infty}\ p(x,y)\ dy\\
&=&2\ \int_{0}^{\infty}\ p(x,y)\ dy
\end{eqnarray}
Now for $x\gg1$, we have: $p(x,y)\sim \frac{A}{(x^2+y^2)^\frac{\gamma +2}{2}}$. Consequently, setting $y=ux$ and considering cases when $x \gg 1$:
\begin{eqnarray}
p_X(x)&\sim & 2\ \int_{0}^{\infty}\ \frac{A}{(x^2+y^2)^\frac{\gamma +2}{2}}\ dy\\
&\sim & 2Ax\ \int_{0}^{\infty}\ \frac{1}{x^{\gamma + 2}(1+u^2)^\frac{\gamma +2}{2}}\ du\\
&\sim & \frac{A\sqrt{\pi}\Gamma  \left( \frac{\gamma +1}{2}\right)}{x^{\gamma + 1}\Gamma  \left( \frac{\gamma}{2}+1\right)}\\
&\sim & \frac{C}{x^{\gamma + 1}}
\end{eqnarray}
where we have set $C=A\sqrt{\pi}\frac{\Gamma  \left( \frac{\gamma +1}{2}\right)}{\Gamma  \left( \frac{\gamma}{2}+1\right)}$.
\medskip

\noindent Now we can express $F_N(x)$ for large $x$ and large $n$:
\begin{eqnarray}
F_N(x)&\sim &\left[ 1-\int _x ^\infty\ \frac{C}{x'^{\gamma + 1}}\ dx'\right]^N\nonumber\\
&\sim &\left[ 1-\frac{C}{\gamma\ x^{\gamma}}\right]^N\nonumber\\
&\sim & e^{-\frac{NC}{\gamma\ x^{\gamma}}}\label{Fnc}
\end{eqnarray}
Whence, still for large $x$ and large $N$:
\begin{equation}
F_N'(x)\sim \frac{NC}{x^{\gamma + 1}}\ e^{-\frac{NC}{\gamma\ x^{\gamma}}}\label{Fn'c}
\end{equation}
\medskip

\noindent We now insert (\ref{Fn'c}) into (\ref{Ia2}), setting 
$u=\frac{NC}{\gamma\ x^{\gamma}}$:
\begin{eqnarray}
I_N^{(2)} &\sim &\int_{0}^{\infty}\ 
\left(\frac{NC}{\gamma}\right)^\frac{2}{\gamma}\ 
u^{-\frac{2}{\gamma}}\ e^{-u}\, du \nonumber\\
&\sim & \left(\frac{NC}{\gamma}\right)^\frac{2}{\gamma}\ \Gamma \left( 1-\frac{2}{\gamma}\right)\label{In2f}
\end{eqnarray}
\medskip

\noindent Let us now examine the second term in equation (\ref{AIs2}). As before, we focus on the large-$x$ part of the integral, which will dominate. We rewrite it, so as to bring it down 
to the calculation that we have done in the previous paragraph:
\begin{eqnarray}
&&N\int _K^\infty\int _{-\infty}^\infty\ y^2\ p(x,y)\ F_{N-1}(x)\ dx\ dy\nonumber\\
&=&\int _K^\infty\ F_N'(x)\ \frac{\int _{-\infty}^\infty\ y^2\ p(x,y)\ dy}{p_X(x)}\ dx\nonumber\\
&\sim & \int _K^\infty\ F_N'(x)\ x^2\ \frac{A}{C}\ \int _{-\infty}^\infty\ \frac{u^2}{(1+u^2)^{\frac{\gamma +2}{2}}}\ du\ dx\nonumber\\
&\sim &  \frac{1}{\gamma - 1}\ \int _K^\infty\ F_N'(x)\ x^2\ dx
\end{eqnarray}
This last integral is, up to the factor $\frac{1}{\gamma - 1} $, the same as (\ref{Ia2}); consequently, we obtain:
\begin{equation}
\langle A_N \rangle \mathop{\sim}_{N\rightarrow \infty} \left( 1-\frac{1}{\gamma - 1}\right)\ \int _K^\infty\ x^2\ F_N'(x)\ dx,\nonumber
\end{equation}
which, combined to (\ref{In2f}) and simplified, yields:
\begin{equation}
\langle A_N \rangle \mathop{\sim}_{N\rightarrow \infty} \frac{\gamma}{\gamma - 1}\ \left(\frac{C}{\gamma}\right)^{\frac{2}{\gamma}}\Gamma \left( 2\left( 1-\frac{1}{\gamma}\right)\right)\ 
N^{\frac{2}{\gamma}}.
\end{equation}
This coincides with the result of Carnal~\cite{Ca}. 
\end{quotation}

\bigskip

{\bf {Asymptotic results for general isotropic case:}}

\medskip

The large $N$ asymptotic analysis for a general isotropic distribution, 
both for the 
mean perimeter and the mean area, can be done following
the details presented in the above two examples. We just provide
a summary here without repeating the details.
\medskip

\textbf{Average Perimeter:}

\begin{itemize}

\item \textit{Exponential tails} ($p(x,y) \sim A e^{-(x^2+y^2)^{\alpha/2}}$ when $(x^2+y^2)\rightarrow \infty$) $$\langle L_{N} \rangle \sim 2\pi \log ^{1/\alpha} N$$

\item \textit{Power-law tails} ($p(x,y) \sim \frac{A}{(x^2+y^2)^{(\frac{\gamma + 1}{2})}}$): $$\langle L_{N} \rangle \sim 2\pi 
\left(\dfrac{A\,B(\frac{1}{2},\frac{\gamma+1}{2})}{\gamma}\right)^{\frac{1}{\gamma}}\Gamma\left(1-\frac{1}{\gamma}\right)\,N^{\frac{1}{\gamma}},$$ 
where $B(x,y)$ is the beta function.

\item \textit{Truncated tails} ($p(x,y) \sim A(a-\sqrt{x^2+y^2})^{\gamma -1}$): $$\langle L_{N} \rangle \sim 2\pi \left( a-\dfrac{f(a,\gamma)}{N^{\frac{2}{2\gamma+1}}}\right),$$ with 
$$f(a,\gamma)=\dfrac{(\gamma+\frac{1}{2})^{\frac{1-2\gamma}{1+2\gamma}}\Gamma(\frac{2}{1+2\gamma})}{(2A\sqrt{2a})^{\frac{2}{1+2\gamma}}}$$
\end{itemize}
\medskip

Therefore, we do find as expected the distinction between the three different universality classes of extreme-value statistics. The sets of independent points drawn from distributions with 
exponential tails have, on average when $N$ becomes large, a convex hull whose perimeter increases more slowly (in powers of $\log N$) than sets drawn from distribution with power-law tails 
(for which the growth of the perimeter is in powers of $N$), which reveals the lesser probability of having points very far from the origin in exponential-tailed distributions than in 
power-law tailed distributions.
\medskip

\textbf{Average area:}

\begin{itemize}

\item \textit{Exponential tails}: $$ \langle A_N \rangle \sim \pi \log ^{\frac{2}{\alpha}}N $$

\item \textit{Power-law tails}: $$ \langle A_N \rangle \sim \pi (\frac{\gamma}{\gamma-1}) \left(\dfrac{AB(\frac{1}{2},\frac{\gamma+1}{2})}{\gamma}\right)^{\frac{2}{\gamma}} \Gamma 
\left(2-\frac{2}{\gamma}\right) N^{\frac{2}{\gamma}} $$

\item \textit{Truncated tails} : $$\langle A_N \rangle \sim \pi a^2 \left (1-\dfrac{8\ f(a,\gamma)}{3\ N^{\frac{2}{2\gamma+1}}}\right),$$ with 
$$f(a,\gamma)=\dfrac{(\gamma+\frac{1}{2})^{\frac{1-2\gamma}{1+2\gamma}}\Gamma(\frac{2}{1+2\gamma})}{(2A\sqrt{2a})^{\frac{2}{1+2\gamma}}}$$
\end{itemize}
\medskip

We find for the area the same characteristics as for the perimeter as far as the relative growths of convex hulls are concerned, depending on the shape of the initial distribution of the 
points. It is particularly worth noting that, in the case of exponential-tailed distributions, the asymptotic behaviour of the average perimeter and area of the convex hull correspond to 
the geometrical quantities of a circle centered on the origin and with radius $\log ^{\frac{1}{\alpha}}N$ ($\alpha$ being the characteristic exponent of the initial distribution's 
exponential tails).\footnote{This is to be compared with Geffroy's results \cite{Ge1}, p.~\pageref{Geffr}.}

\subsection{A non-isotropic case: points distributed uniformly in a square}

Let us now examine a non-isotropic case: computing the average perimeter of the 
convex hull of $N$ independent points distributed uniformly in a 
square of side $a$. The bivariate 
probability density of the sample can be written as:
\begin{equation}
p(x,y)= \frac{1}{a^2}\Theta(\frac{a^2}{4}-x^2)\Theta(\frac{a^2}{4}-y^2),
\end{equation}
where $\Theta$ is the Heaviside step function.
\medskip

We will consider, as described in the introductory part of this section, the projection of the sample on the line through the origin making an angle $\theta$ with the $x$-axis. We write the 
random variable corresponding to the projection of a sample point $z\equiv x\cos \theta + y\sin \theta$. Its density will be given by :
\begin{equation}
q(z)=\frac{1}{a^2}\iint_{-\frac{a}{2}}^{\frac{a}{2}}\ \delta (z-x\cos \theta -y\sin \theta)\Theta(\frac{a^2}{4}-x^2)\Theta(\frac{a^2}{4}-y^2)\ dx\ dy,\label{qz}
\end{equation}
where $\delta$ is the Dirac delta function.
\medskip

Using the symmetry of the square, we can focus on $0\leq\theta\leq\frac{\pi}{4}$ and write $$x=\frac{z-y\sin \theta}{\cos \theta}.$$ This enables us to simplify (\ref{qz}):
\begin{equation}
 q(z)=\frac{1}{a^2\cos \theta}\int_{-\frac{a}{2}}^{\frac{a}{2}}\Theta(\frac{a^2}{4}-\left( \frac{z-y\sin \theta}{\cos \theta}\right)^2)\Theta(\frac{a^2}{4}-y^2)\ dy.
\end{equation}
Enforcing the condition that any point of the sample lies inside the square and thus making the Heaviside functions non-zero, we have:
\begin{eqnarray}
\text{(i)}&-\frac{a}{2}\leq y \leq \frac{a}{2}&\\
\text{(ii)}&-\frac{a}{2}\leq \frac{z-y\sin \theta}{\cos \theta} \leq \frac{a}{2}&\\
\therefore & \max \left( -\frac{a}{2},\frac{z-\frac{a}{2}\cos \theta}{\sin \theta}\right)\leq y \leq \min \left( \frac{a}{2},\frac{z+\frac{a}{2}\cos \theta}{\sin \theta}\right)&
\end{eqnarray}
and:
\begin{equation}
-\frac{a}{2}(\cos\theta +\sin \theta)\leq z \leq \frac{a}{2}(\cos\theta +\sin \theta)
\end{equation}
There will be 3 cases:
\begin{enumerate}

\item For $$ -\frac{a}{2}(\cos\theta +\sin \theta)\leq z\leq \frac{a}{2}(\sin \theta - \cos \theta),$$ the $y$-coordinate will vary between $-\frac{a}{2}$ and $\frac{z+\frac{a}{2}\cos 
\theta}{\sin \theta} $ and therefore:
\begin{equation}
q(z)=\frac{z+\frac{a}{2}(\cos \theta + \sin \theta)}{a^2\cos \theta \sin \theta}.\label{qz1}
\end{equation}

\item For $$\frac{a}{2}(\sin \theta - \cos \theta)\leq z\leq \frac{a}{2}(\cos \theta - \sin \theta),$$ the $y$-coordinate will vary between $-\frac{a}{2}$ and $\frac{a}{2}$ and therefore:
\begin{equation}
q(z)= \frac{1}{a\cos \theta}.\label{qz2}
\end{equation}

\item For $$\frac{a}{2}(\cos\theta -\sin \theta)\leq z\leq \frac{a}{2}(\cos \theta + \sin \theta)$$ the $y$-coordinate will vary between $\frac{z-\frac{a}{2}\cos \theta}{\sin \theta} $ and 
$\frac{a}{2}$  and therefore:
\begin{equation}
q(z)=\frac{\frac{a}{2}(\cos \theta + \sin \theta)-z}{a^2\cos \theta \sin \theta}.\label{qz3}
\end{equation}
\end{enumerate}

\noindent To lighten the notation, let us write henceforth:
\begin{eqnarray}
a_\theta &=& \frac{a}{2}(\cos \theta+\sin \theta)\\
b_\theta &=& a^2 \cos \theta \sin \theta
\end{eqnarray}

Denoting as before by $M_N(\theta)$ the value of the support function of the sample at angle $\theta$, that is, the value of the maximal projection on direction $\theta$, we have:
\begin{eqnarray}
\langle M_N(\theta) \rangle &=& \int_{-a_\theta}^{a_\theta}\ z\ F'_{\theta,N}(z)\ dz\\
&=& \left[zF'_{\theta,N}(z)\right]_{-a_\theta}^{a_\theta}- \int_{-a_\theta}^{a_\theta}\ F_{\theta,N}(z)\ dz\\
&=&a_\theta - I_\theta,
\end{eqnarray}
where:
\begin{eqnarray}
F_{\theta,N}(z)&=& \left[\int_{-a_\theta}^z\ q(z')\ dz'\right]^N\\
I_\theta &=& \int_{-a_\theta}^{a_\theta}\ F_{\theta,N}(z)\ dz
\end{eqnarray}

To compute $I_\theta$ we make use of our knowledge of $q(z)$ (Eqs.~(\ref{qz1}), 
(\ref{qz2}), (\ref{qz3}))  and we obtain:
\begin{multline}
\langle M_N(\theta) \rangle = a_\theta - \frac{\sin \theta \tan ^N \theta}{2^{N-1}(2N+1)}-\frac{\cos \theta}{2^N(N+1)}\left[(2-\tan \theta)^{N+1}- \tan ^{N+1}\theta\right] \\- 
\sqrt{b_\theta \tan \theta}\ {}_2F_1\left( \frac{1}{2},-N;\frac{3}{2};\frac{\tan \theta}{2}\right),
\end{multline}
${}_2F_1$ being a hypergeometric function.
\medskip

\noindent Using known facts about the asymptotic behaviour of hypergeometric series \cite{AbSteg}, $\langle M_N(\theta) \rangle$ can be seen to behave in the following way for large $N$:
\begin{equation}
\langle M_N(\theta) \rangle \sim a_\theta - \sqrt{\frac{\pi b_\theta}{2N}}+o\left( \frac{1}{\sqrt{N}}\right)
\end{equation}
This then yields the desired result:
\begin{eqnarray}
\langle L_{N} \rangle &=& 8\int_0^\frac{\pi}{4}\ \langle M_N (\theta)\rangle\\
&\sim & 4\ a \left(1-\pi \frac{\Gamma \left(\frac{3}{4}\right)}{\Gamma \left(\frac{1}{4}\right)\ \sqrt{N}}\right).
\end{eqnarray}
This is the same as Rényi and Sulanke's \cite{RS2}, which they obtained from a different approach. Note that, as in the case of points distributed uniformly inside a disk, the average 
perimeter of the convex hull tends to that of the boundary of the support ---~here $4a$, the perimeter of the square~--- when the number $N$ of points becomes large. However, the 
convergence here is slower than for a support with a smooth boundary like the disk: $N^{-\frac{1}{2}}$ versus $N^{-\frac{2}{3}}$. One can think that, physically and statistically, it is 
somehow "more difficult" for the points of the sample to reach inside the corners of the square, making the convergence of the convex hull towards the square all the more slower.

\section{Correlated Points: One or more Brownian Motions}

As mentioned before, one of the advantages of the support function 
approach that we use in this paper 
is its generality: it can be applied to samples with correlations 
as 
well as to samples of independent points. In this Section, we study, using this 
method, the convex hull of $n$ planar Brownian paths, 
a topic that has so far been considered only in the $n=1$ 
case \cite{Lev, El,Ta,  Go, Go2}.

Beyond its interest from a theoretical point of view, the study of the convex hull of $n$ planar Brownian paths can be motivated by a question of particular relevance to the conservation of 
animal species in their habitat, as we shall see before giving the details of results.

\subsection{Planar Brownian paths and home-range\label{ApplEco}}
A question that ecologists often face, in particular in
designing a conservation area to preserve a given
animal population \cite{ConsPlan}, is how to
estimate the home-range of this animal population.
Roughly speaking this means the following.
In order to survive over a certain length of time, the animals need
to search for food and hence explore a certain region of space. How much space
one needs to assign for a group of say $n$ animals?  
For 
instance, in the case of species having a 
nest to which they return, say, every night, the "length of time" is just
the duration of a day. In ecology, the home range is simply 
defined as the territory explored by the herd during 
its daily search for food over a fixed length of time. Different methods are used 
to estimate this territory, 
based on the monitoring of the animals' positions \cite{Worton,Luca}. 
One of these consists in simply the 
minimum convex polygon 
enclosing all monitored positions, called the convex hull. While this may seem 
simple minded, it 
remains, under certain circumstances, the best way to proceed~\cite{Folia}.

\begin{figure}[h]
\begin{center}
\includegraphics[height=3cm, width=5cm]{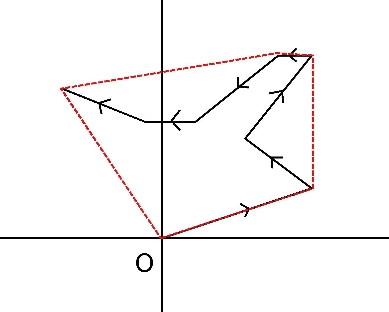}
\end{center}
\caption{Convex hull of a 7-step random walk}\label{RW}
\end{figure}

\begin{figure}[h]
\begin{center}
\includegraphics[height=3cm, width=5cm]{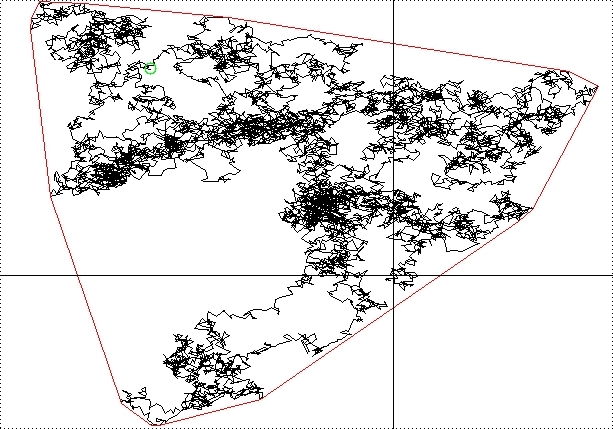}
\end{center}
\caption{Convex hull of planar Brownian motion}\label{BrCH}
\end{figure}

The monitored 
positions, for one animal, will appear as the 
vertices of a path whose statistical properties will depend on the type 
of motion the animal is performing. In particular, during phases of food 
searching known as foraging, the monitored positions can be described as 
the vertices of a random walk in the plane~\cite{ERW,BRW,EKRW}. For 
animals whose daily motion consists mainly in foraging, quantities of 
interest about their home range, such as its perimeter and area, can be 
estimated through the average perimeter and area of the convex hull of the 
corresponding random walk (Fig.~\ref{RW}).
If the recorded positions are 
numerous (which might result from a very fine and/or long monitoring), the 
number of steps of the random walker becomes large and
to a good approximation the trajectory of a discrete-time planar random walk 
(with finite variance of the step sizes) can
be replaced by a continuous-time planar Brownian motion of a certain duration $T$. (fig.~\ref{BrCH}).

\begin{figure}[h]
\begin{center}
\includegraphics[height=4cm, width=7cm]{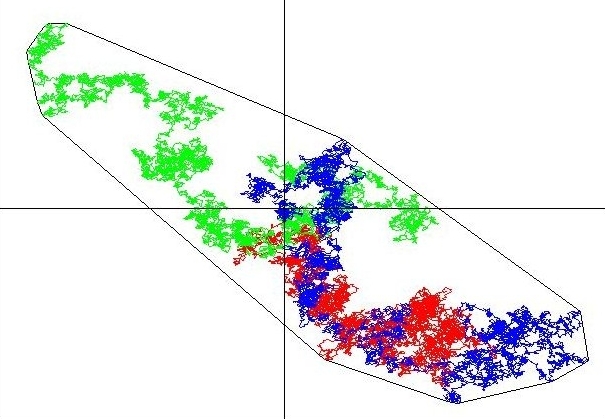}
\end{center}
\caption{Convex hull of 3 independent, closed Brownian paths, starting at the origin $O$.}\label{NBrCH}
\end{figure}

The home range of a single animal can thus be characterized by
the mean perimeter and area of the convex hull of a planar Brownian motion 
of duration $T$ starting at origin $O$.
Both `open' (where the endpoint of the path
is free) and `closed' paths (that are constrained to
return to the origin in time $T$) are of interest. The latter corresponds, for instance,
to an animal returning
every night to its nest after spending the day foraging in the
surroundings. As we have seen in our review of existing results, the average perimeter and area of the convex hull of an open Brownian path are known \cite{Ta,El}, as is the average 
perimeter for a closed path \cite{Go}. It seems natural and logical to seek an extension of these results to an arbitrary number of paths (fig.~\ref{NBrCH}), both from a theoretical point 
of view and from an ecological one, since many animals live in herds. We show first how to use the support-function method for $n=1$ planar Brownian paths and then for $n>1$.

\subsection{Convex hull of a planar Brownian path}

We consider here a planar Brownian path of duration $T$, starting from the origin~$O$: $$\mathcal{B}(\tau)=(x(\tau),y(\tau))$$ with $$0\leq \tau\leq T,$$ $x(\tau)$ and $y(\tau)$ being 
standard 1-dimensional Brownian motions of duration $T$ obeying the following Langevin equations: $${\dot x}(\tau)= \eta_x(\tau)$$
and $${\dot y}(\tau)=\eta_y(\tau)$$ where $\eta_x(\tau)$ and $\eta_y(\tau)$
are independent Gaussian white noises, with zero mean and \textit{delta}-correlation:  $$\langle \eta_{.}(\tau)\eta_{.}(\tau')\rangle 
=\delta(\tau-\tau').$$ Let us note incidentally that this implies: $$\langle x^2(\tau)\rangle=\tau$$
and \begin{equation}\langle y^2(\tau)\rangle =\tau\label{ysq}\end{equation}. 

\begin{figure}
\begin{center}
\subfigure[]{\includegraphics[height=3.6cm, width=4.25cm]{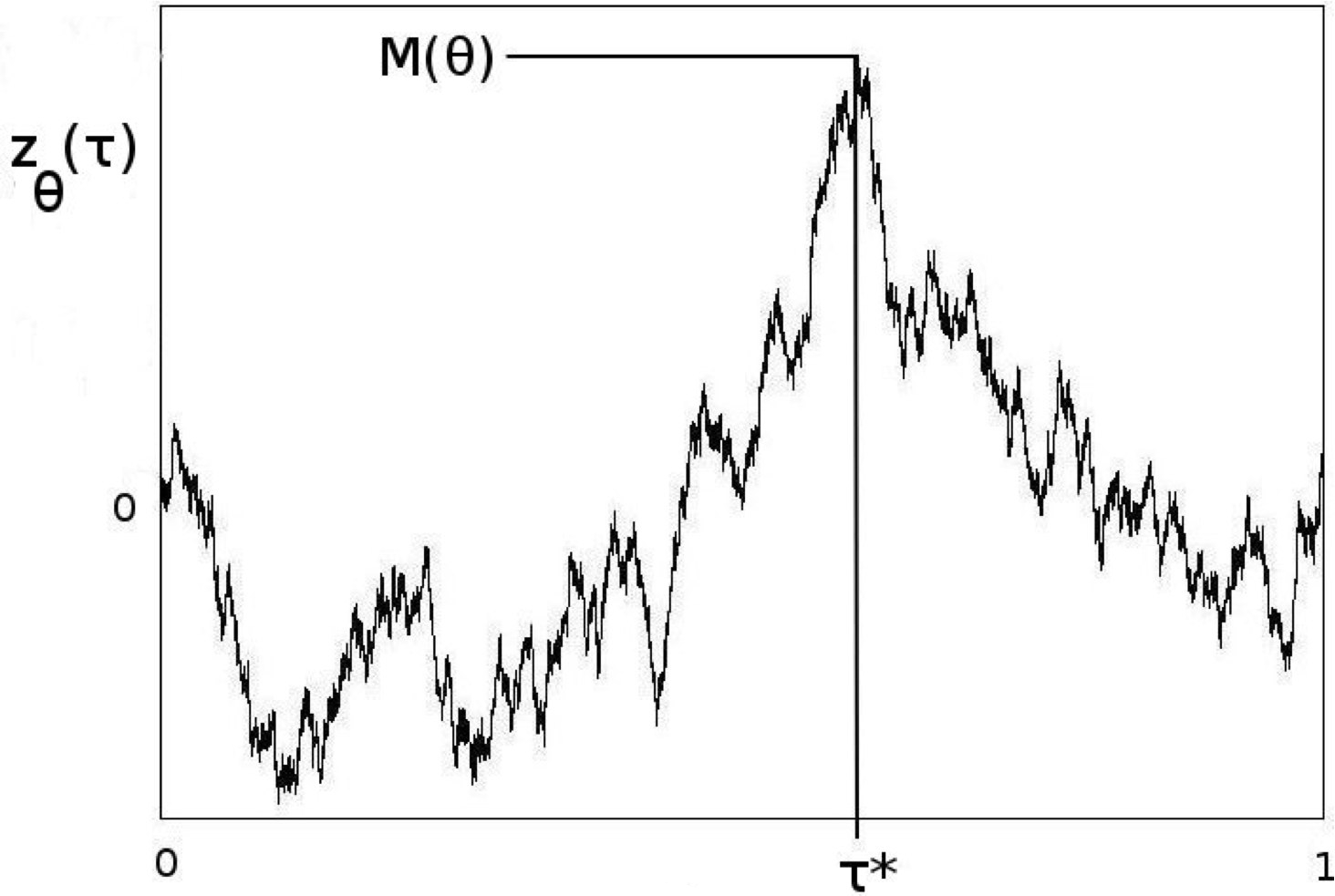}\label{z}}
\subfigure[]{\includegraphics[height=3.6cm, width=4.25cm]{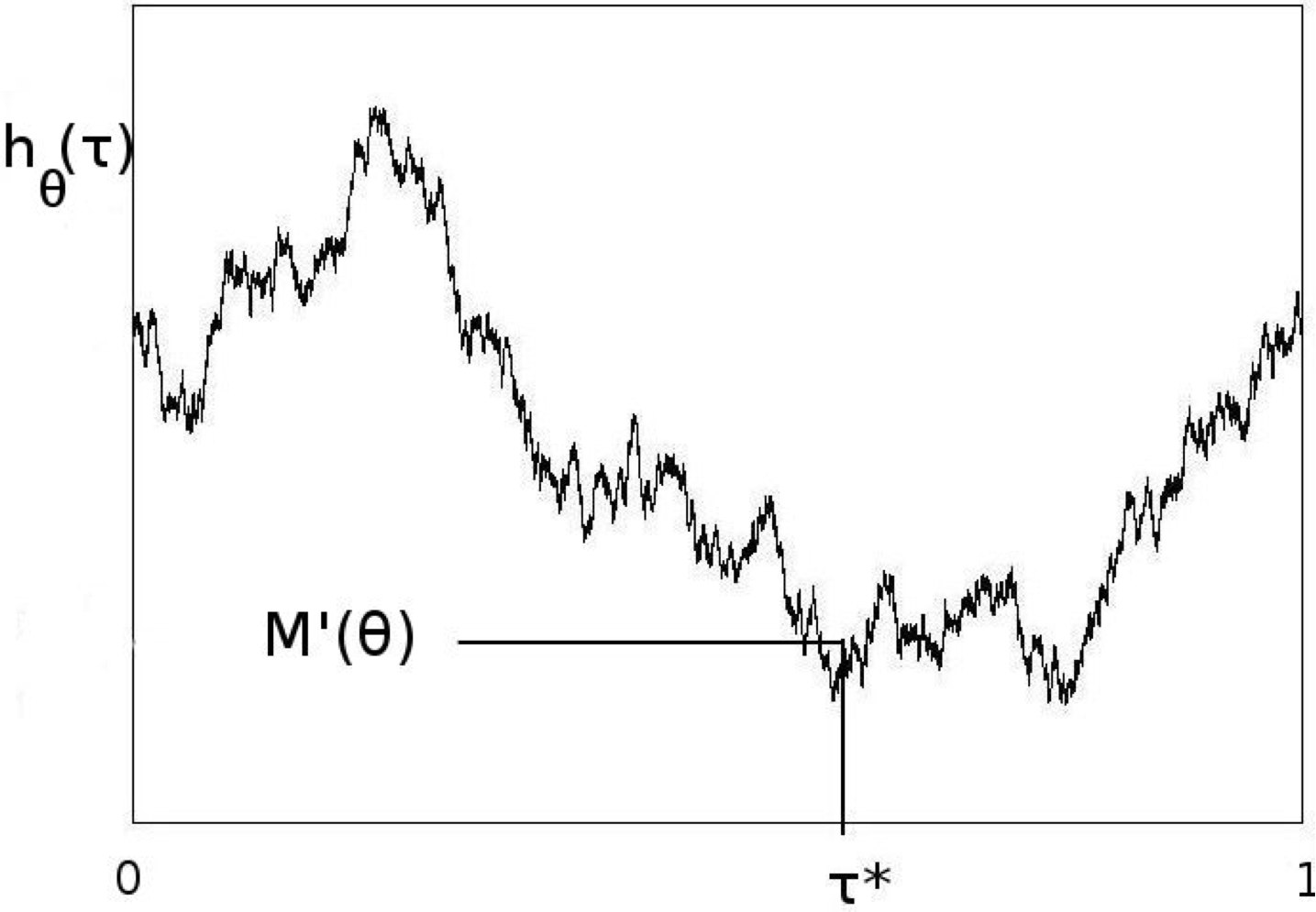}\label{h}}
\caption{(a) Time $\tau^*$ when the maximum $M(\theta)$ of $z_\theta(\tau)$  is attained and (b) 
corresponding value $M'(\theta)=h_\theta(\tau^*)$}
\end{center}
\end{figure}

Fix a direction  $\theta$. We use as before (Eqs.~(\ref{zeq}) and (\ref{heq})) the 
projection on direction $\theta$:
$$z_\theta(\tau)= x(\tau)\cos \theta+y(\tau) \sin \theta$$ and
$$h_\theta(\tau)= -x(\tau) \sin \theta+y(\tau)\cos\theta.$$
Now, $z_\theta$ and $h_\theta$ are two independent 1-dimensional Brownian motion (each of duration $T$),
parametrized by $\theta$.  
It thus appears that $M(\theta)$ is simply the maximum of the 1-dimensional Brownian motion $z_{\theta}(\tau)$ on the interval $\tau\in [0,T]$, i.e., $$M(\theta)= \mathop{\max}_{\tau\in 
[0,T]}[ z_{\theta}(\tau)].$$  
Furthermore, if we write $\tau^*$ the time at which this maximum is attained, then: $$M(\theta)= z_\theta(\tau^*)= x(\tau^*) \cos \theta+ y(\tau^*) \sin \theta.$$
Deriving with respect to $\theta$ gives:  
$$M'(\theta)= -x(\tau^*) \sin \theta+ y(\tau^*) \cos \theta= h_{\theta}(\tau^*).$$
In words, if $M(\theta)$ is the maximum of the first Brownian motion
$z_{\theta}(\tau)$, $M'(\theta)$ corresponds to the value of the second, independent motion
$h_\theta(\tau)$ at the time $\tau=\tau^*$ when the first one attains its maximum.
(\textit{cf.}  Fig.~\ref{z} and \ref{h}).

In particular, when $\theta=0$, $z_0(\tau)=x(\tau)$
and $h_0(\tau)= y(\tau)$, and $M(0)$ is then the maximum of $x(\tau)$ on the interval $\tau\in [0,T]$
while $M'(0)=y(\tau^*)$ is the value of $y$ at the time $\tau^*$ when $x$ attains its maximum. 
\medskip

Recall that in isotropic cases, Cauchy's formulae (Eqs.~(\ref{AC1}) and (\ref{AC2})) 
simplify to:
\begin{eqnarray}
\langle L\rangle &=& 2\pi \ \langle M(0) \rangle \label{IAC1}\\
\langle A\rangle &=& \pi \left( \langle [M(0)]^2 \rangle - \langle
[M'(0)]^2\rangle\right). \label{IAC2}
\end{eqnarray}
The planar motion that we are considering here is assumed to be isotropic and we will thus use this version of the formulae.
\medskip

The distribution of the maximum of a 1-dimensional Brownian motion $x(\tau)$ on $[0,T]$ is known, and given by the cumulative distribution function: \begin{equation}F(M)={\rm Prob}[M(0)\le 
M]=\ \text{erf}\left(\frac{M}{\sqrt{2T}}\right),\label{Q1M}\end{equation} with
$${\rm erf}(z)= \frac{2}{\sqrt{\pi}}\,\int_0^z e^{-u^2}\, du.$$
The first two moments of this distribution are readily computed:
$$\langle M(0)\rangle = 
\sqrt{\frac{2T}{\pi}}$$ and $$\langle [M(0)]^2\rangle = T.$$
Equation (\ref{IAC1}) then yields the average perimeter of the convex hull of a planar Brownian path, 
\begin{equation}
\langle L\rangle=\sqrt{8\pi T}.\label{LFM}
\end{equation}
\medskip

It is slightly more complex to compute the average area enclosed by the convex hull as one then needs to compute $\langle 
[M'(0)]^2\rangle$.  
Let us first recall (Eq.~(\ref{ysq})) that for a given $\tau^*$, $$E [y^2(\tau^*)] = 
\tau^*$$
(since $y$ is a standard Brownian motion), the expectation being taken over all possible realisations of $y$ at fixed $\tau^*$. But $\tau^*$ is itself a random variable, since it is the 
time at which the first process, $x$, attains its maximum. 
One therefore also has to average over the probabability density
of 
$\tau^*$ (which  is given by L\'evy's celebrated arcsine law:
$\rho_1(\tau^*) = [\tau^* (T-\tau^*)]^{-1/2}/\pi$); this leads to:
$$\langle [M'(0)]^2\rangle= \langle \tau^*\rangle = T/2;$$ whence one obtains, \textit{via} equation~(\ref{IAC2}), the exact expression for the average area enclosed by the convex hull of 
the motion:
\begin{equation}
\langle A \rangle = \frac{\pi T}{2}.\label{AFM}
\end{equation}
\bigskip

If we now consider a closed Brownian path in the plane, that is, one constrained to return to the origin after time $T$, the reasoning is completely similar, but for $x(\tau)$ and $y(\tau)$ 
which are now Brownian bridges of duration $T$: both start from the origin and are constrained to return to it at time $T$: $$x(0)=x(T)=0$$ $$y(0)=y(T)=0.$$ The distribution of the maximum 
of a Brownian bridge is also known, and its first two moments are given by: $$\langle M(0)\rangle = \sqrt{\frac{\pi T}{8}}$$ and $$\langle [M(0)]^2\rangle =\frac{T}{2}.$$ 
Equation~(\ref{IAC1}) gives us as before the average perimeter of the convex hull: 
\begin{equation}
\langle L\rangle= \sqrt{\frac{\pi^3 T}{2}}.\label{LBM}
\end{equation}

\begin{sloppypar}
To compute the average area enclosed by the convex hull $y(\tau)$, let us first note that for a Brownian bridge, at a fixed time $\tau^*$: $$E [y^2(\tau^*)] = 
\frac{\tau^*(T-\tau^*)}{T}.$$ Let us then recall another well-known 
result: the probability density of $\tau^*$ is uniform. Thus, averaging on 
$\tau^*$, with uniform distribution 
$\rho_1(\tau^*)=1/T$, we obtain: 
$$\langle M'(0)^2\rangle= \langle y^2(\tau^*)\rangle = \frac{T}{6}.$$ 
Finally, as before, equation~(\ref{IAC2}) leads us to an exact expession 
for the average area enclosed by the convex 
hull of a 2d Brownian bridge: 
\begin{equation}
\langle A\rangle =  \frac{\pi T}{3}.\label{ABM}
\end{equation}
\end{sloppypar}
\bigskip

Results (\ref{LFM}) and (\ref{AFM}) had been computed by M. El Bachir \cite{El}, using the same approach as here, hinted at by L. Tak\'{a}cs \cite{Ta}. Equation (\ref{LBM}) is given by A. 
Goldman. The last result, (\ref{ABM}), is, to the best of our knowledge, new, as are those in the next paragraph, regarding the convex hull of 
several planar Brownian motions.

\subsection{Convex hull of $n$ planar Brownian paths}
As mentioned earlier, the method can then be generalized to $n$ independent
planar Brownian paths, open or closed. 
We now have two sets of
$n$ Brownian paths: $x_j(\tau)$ and $y_j(\tau)$ ($j=1,2,\ldots,n$).
All paths are independent of each other. 
Since isotropy holds,
we can still use Eqs. (\ref{IAC1}) and (\ref{IAC2}), except
that $M(0)$ now denotes the global maximum of a set of $n$
independent one dimensional Brownian paths (or bridges for closed paths)
$x_j(\tau)$ ($j=1,2,\ldots,n$), each of duration $T$, 
\begin{equation}
M(0)= \max_{1\le j\le n}\,\max_{0\le \tau\le T}\left[x_1(\tau),x_2(\tau),\ldots, 
x_n(\tau)\right].
\label{MN1}
\end{equation}
Let $j_*$ and $\tau^*$ denote the label of the path and the time at which this
global maximum is achieved. Then, using argument similar to the $n=1$ case, it is easy to 
see that $M'(0)= y_{j_*}(\tau^*)$, i.e., the position of the $j_*$-th $y$ path
at the time when the $x$ paths achieve their global maximum.

To compute the first two moments of $M(0)$, we first compute the distribution
$P_n[M(0),T]$ of the global maximum of $n$ independent Brownian paths (or 
bridges) $x_j(\tau)$. This is a standard extreme value calculation.

\subsection{Open paths}
Consider first $n$ open Brownian paths.
It is easier to compute the cumulative probability, $$F_n(M)= {\rm Prob}[M(0)\le 
M].$$ Since the Brownian paths are independent, it follows that $$F_n(M)= [F(M)]^n,$$
where $$F(M)= {\rm erf}\left(\frac{M}{\sqrt{2T}}\right)$$ for a single path mentioned before.

\begin{sloppypar}
\noindent Knowing this cumulative distribution
$F_n(M(0))$, the first two moments $\langle M(0)\rangle$ and $\langle [M(0)]^2\rangle$
can be computed for all $n$. Using the result for $\langle M(0)\rangle$ in Eq. 
(\ref{IAC1}) gives us the mean perimeter, $\langle L_N\rangle =\alpha_n \sqrt{T}$
with
\begin{equation}
\alpha_N = 4 n \sqrt{2\pi} \int _0^\infty du\ u\ e^{-u^2}
\left[\text{erf}\left(u\right)\right]^{n-1}.
\label{R1}  
\end{equation}
The first few values are:
\begin{eqnarray*}
\alpha_1 &=& \sqrt{8\pi}=5.013..,\\
\alpha_2 &=& 
4\sqrt{\pi}=7.089..\\
\alpha_3 &=& 24\,\frac{{\rm tan}^{-1}\left(1/\sqrt{2}\right)}{\sqrt{\pi}}=8.333..
\end{eqnarray*}
 (see Fig. \ref{plot} for
a plot of $\alpha_n$ vs. $n$).\\
For large $n$, one can
analyse the integral in Eq.~(\ref{R1}) by the saddle point method giving:
\begin{equation}
\alpha_n \sim 2\pi \sqrt{2 \log n}.\label{alphan}
\end{equation}
(Details of the analysis are given in appendix \ref{AsBe})\\
This logarithmic dependence on $n$
is thus a direct consequence of extreme value statistics~\cite{EVSG2}
and the calculation of the mean perimeter
of the convex hull of $n$ paths is a nice application of the extreme value 
statistics.
\end{sloppypar}
\medskip

To compute the mean area, we need to calculate $\langle [M'(0)]^2\rangle$ in
Eq.~(\ref{IAC2}). We proceed as in the $n=1$ case. For a fixed label $j$ and fixed 
time $\tau$: $$E[y_j^2(\tau)]=\tau,$$ which follows
from the fact that $y_j(\tau)$ is simply a Brownian motion.
Thus: $$E[y_{j_*}^2(\tau^*)]=\tau^*.$$

\begin{figure}
\begin{center}
\includegraphics[height=5.5cm,width=8cm]{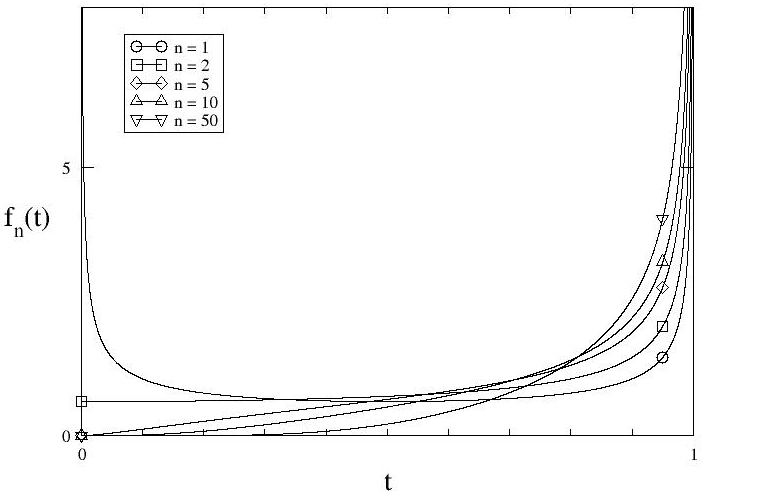}
\caption{Probability density $f_n(t)$ of the time $t$ at which the global 
maximum of $n$ Brownian motions, each of unit duration, is attained, as
given by the formula in Eq. (\ref{maxtN}).
}
\end{center}
\label{DTMB}
\end{figure}

\noindent Next, we need to average over
$\tau^*$ which is the time at which the global maximum in Eq. (\ref{MN1})
happens. The probability density ${\rho}_n(\tau^*)$ of the time $\tau^*$ of the 
global
maximum of $n$ independent Brownian motions (each of duration $T$), to our 
knowledge,
is not known in the probability literature. We were able to compute
this exactly for all $n$ (details are given in appendix~\ref{ToMN}). We find
that $$\rho_n(\tau^*)= \frac{1}{T}\,f_n(\tau^*/T)$$ where the scaling function
$f_n(z)$ is given by 
\begin{equation}
f_n(z)= \frac{2n}{\pi\sqrt{z(1-z)}}\,\int_0^{\infty}dx\, x\, e^{-x^2}\, 
\left[{\rm erf}\left(x\sqrt{z}\right)\right]^{n-1}.
\label{maxtN}
\end{equation}  
A plot of $f_n(z)$ for various values of $n$ is given in Fig. (\ref{DTMB}).
\bigskip

It is easy to check that for $n=1$, it reproduces the arcsine law mentioned before.\\
Averaging over $\tau^*$ drawn from this distribution, we can then
compute 
$$\langle [M'(0)]^2\rangle = \int_0^{T} \tau^* \rho_n(\tau^*)\, d\tau^*.$$
Substituting this in Eq. (\ref{IAC2}) gives the exact mean area 
for all $n$, $\langle A_n \rangle =\beta_n T$ with
\begin{equation}
\beta_n= {4n}\,{\sqrt{\pi }}\, \int _0 ^\infty du\ u\
\left[\text{erf}(u)\right]^{n-1}\left(u e^{-u^2}-g(u)\right)
\label{R3}
\end{equation}
where $$g(u)=\frac{1}{2\sqrt{\pi}}\int_0^1 \frac{e^{-{u^2}/{t}}\
dt}{\sqrt{t(1-t)}}.$$\\
For example, the first few values are given by:
\begin{eqnarray}
\beta_1 &=& \pi/2=1.570..\\
\beta_2 &=& \pi= 3.141..\\
\beta_3 &=& \pi+3-\sqrt{3}= 4.409..
\end{eqnarray}
(Fig. \ref{plot} shows a plot of $\beta_n$ vs $n$).\\ The large-$n$ analysis 
(details in appendix \ref{AsBe})
gives: \begin{equation}
\boxed{\beta_n \sim 2\pi \ln n.}\label{bngd}
\end{equation}

\subsection{Closed paths}
\begin{figure}
\begin{center}
\includegraphics[height=5.5cm,width=8cm]{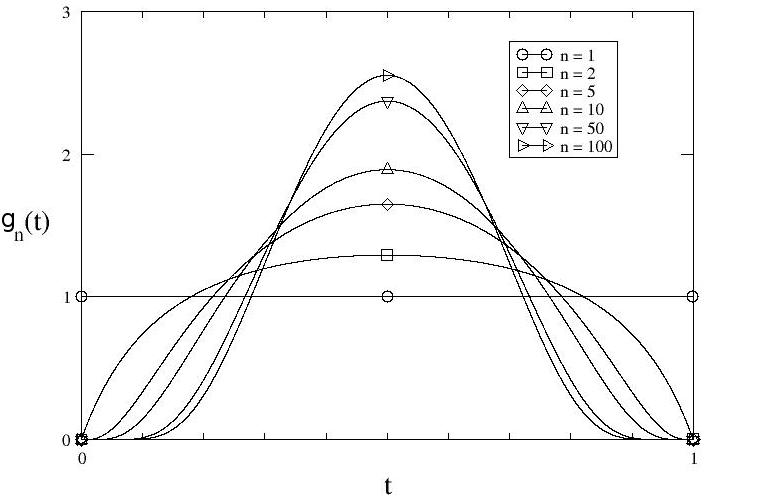}
\caption{Probability density $g_n(t)$ of the time $t$ at which the global 
maximum of $n$ Brownian bridges, each of unit duration, is attained, as
given by the formula in Eq. (\ref{gnz}).}
\label{DTPB}
\end{center}
\end{figure}

For $n$ closed Brownian planar paths one proceeds in a similar way. The differences are:
\begin{itemize}

\item the cumulative distribution function of the maximum of a single 1-dimensional motion is not given by equation~\ref{Q1M} but by:
$$F(M) = 1-e^{-\frac{2M^2}{T}} $$

\item the propagator of the 1-dimensional motion obtained by projection on the $x$-axis is that of a Brownian bridge and so: $$E[y_{j_*}^2(\tau^*)] =\frac{\tau^*(T-\tau^*)}{T}$$

\item the probabilty density of the time $\tau^*$ at which the maximum of $n$ 
1-dimensional 
Brownian bridges occurs is given by (see details in appendix B):  
$$\rho_n(\tau^*) = \frac{1}{T}g_n\left(\frac{\tau^*}{T}\right)$$
where the scaling function $g_n(z)$ is given by
\begin{equation}  
g_n(z)=\frac{4n}{\sqrt{\pi}}\int_0^\infty\ u^2\ 
e^{-u^2}\ \left[1-e^{-4u^2\,z(1-z)}\right]^{n-1}\ du.
\label{gnz}
\end{equation}
\end{itemize}
\medskip
A plot of $g_n(z)$ for different $n$ is given in Fig. (\ref{DTPB}).

\noindent Following then the same route as for open paths, we find that the mean perimeter and area are given by: $$\langle L_n\rangle =\alpha_n(c) 
\sqrt{T}$$ and $$\langle A_n\rangle =\beta_n(c) T$$ where, for all $n$, 
\begin{eqnarray}
\alpha_n(c)&= & \frac{\pi ^{3/2}}{\sqrt{2}}\sum
_{k=1}^{n}\binom{n}{k}\frac{(-1)^{k+1}}{\sqrt{k}} \label{R2} \\
\beta_n(c) &=& \frac{\pi}{2}\left[\sum
_{k=1}^{n}\frac{1}{k}-\frac{n}{3}+\frac{1}{2}\sum_{k=2}^{n}(-1)^k\, f(k)\right]\label{R4}
\end{eqnarray}
and $$f(k)=\binom{n}{k}\,(k-1)^{-3/2}\left(
k\tan^{-1}(\sqrt{k-1})-\sqrt{k-1}\right).$$\\
The first few values are: 
\begin{eqnarray*}
\alpha_1(c)&=& \sqrt{\pi^3/2}=3.937.\\
\alpha_2(c)&=& \sqrt{\pi^3}(\sqrt{2}-1/2)=5.090..\\
\alpha_3(c)&=& 
\sqrt{\pi^3}(3/\sqrt{2}-3/2+1/\sqrt{6})=5.732..
\end{eqnarray*}
and
\begin{eqnarray}
\beta_1(c)&=&\pi/3=1.047..\\
\beta_2(c)&=& \pi(4+3\pi)/24=1.757..\\
\beta_3(c)&=& 2.250..
\end{eqnarray}
 (see Fig. \ref{plot} for a plot of $\alpha_n(c)$
and $\beta_n(c)$ vs. $n$).\\
Large $n$ analysis (details in section \ref{AsBe}) shows that:
\begin{equation}
\boxed{\alpha_n(c)\sim 
\pi \sqrt{2\ln n}}\label{ancgd}
\end{equation} and 
\begin{equation}
\boxed{\beta_n(c) \sim \frac{\pi}{2}
\ln n,}\label{bncgd}
\end{equation}
smaller respectively by a factor $1/2$ and $1/4$ than the corresponding results for open paths ---~as one's intuition might suggest, given that a closed path is enforced to return to 
the origin.

\subsection{Numerical simulations and discussion}

We illustrate our analytical results on the convex hull of planar Brownian motion with some elementary numerical simulations. These require first to generate Brownian paths, and then to 
compute numerically the convex hull of the paths. Here we have used a simple algorithm known as Graham scan \cite{Grah}\footnote{This algorithm is not the quickest one, but our aim was 
mainly illustrative. The question of convex-hull-finding algorithms is a classic one in computer science \cite{Seidel, DevAL}.}.
\begin{figure}[!h]
\begin{center}
\includegraphics[height=7cm, width=9cm]{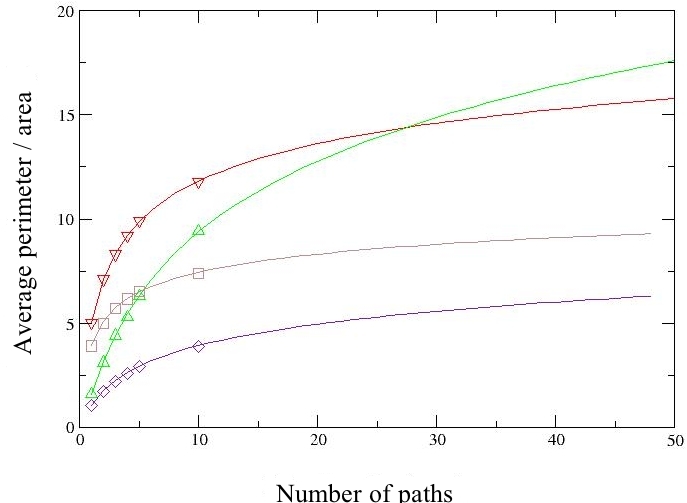}
\caption{Setting $T=1$, the analytical results for average perimeter 
$\alpha_n$ [Eq. 
(\ref{R1}), in red] and area $\beta_n$ [Eq. (\ref{R3}), in green] of $n$ 
open Brownian paths,
and similarly the average perimeter 
$\alpha_n(c)$ [Eq. (\ref{R2}), in brown] and area $\beta_n(c)$ [Eq. 
(\ref{R4}), in purple] of $n$
closed Brownian paths, plotted against $n$. The symbols denote results
from numerical simulations 
(up to $n=10$, with $10^3$ realisations for each point).\label{plot}}
\end{center}
\end{figure}

Let us recall the asymptotic behaviours of the exact formulae plotted in 
figure~\ref{plot}, which are given by equations~(\ref{alphan}), (\ref{bngd}), (\ref{ancgd}) et (\ref{bncgd}).\\
For $n$ open Brownian paths:
\begin{align}
\langle L_n\rangle &\sim 2\pi \sqrt{2 T\ln n }\\
\langle A_n\rangle &\sim 2\pi T\ln n ;
\end{align}
and for $n$ closed Brownian paths:
\begin{align}
\langle L_n^{(c)}\rangle &\sim 
\pi \sqrt{2T\ln n }\\
\langle A_n^{(c)}\rangle &\sim \frac{\pi}{2}
T\ln n .
\end{align}
In both cases, the ratio between the average value of the area of the convex hull and the square of the average value of the perimeter takes, asymptotically, the same value as in the case 
of a circle:
\begin{equation}
\frac{\langle A_n\rangle}{\langle L_n\rangle ^2}
\ \mathop{\simeq}_{n\rightarrow \infty}\ 
\frac{1}{4\pi}.
\end{equation}
Thus, heuristically speaking, for large $n$, the convex hull of 
$n$ planar Brownian paths approaches a circle, centered at
the origin,
whose radius is obtained dividing $\langle L_n\rangle$ by $2\pi$:
\begin{equation}
\boxed{R_n=\sqrt{2T\ln n }}
\end{equation}
for $n$ open paths; and:
\begin{equation}
\boxed{R_n^{(c)}=\sqrt{\frac{T\ln n }{2}}}
\end{equation}
for $n$ closed paths.
\smallskip

Note that for finite $n$ the shape of the convex hull is far from being a
circle. It is only in the $n\to\infty$ limit that it approaches
a circle. Roughly speaking, a large number of trajectories
smoothen their global convex hull into a circular shape.
\begin{figure}[h]
\begin{center}
\includegraphics[height=8cm, width=8cm]{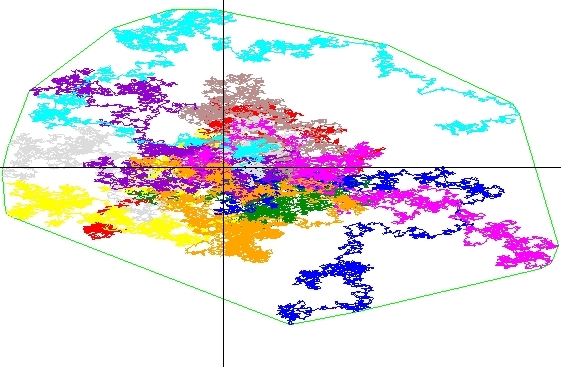}
\end{center}
\caption{Convex hull of 10 open Brownian paths. (Paths are independent and start from the origin.)}\label{10BMHull}
\end{figure}
\bigskip

\begin{figure}[h]
\begin{center}
\includegraphics[height=8cm, width=8cm]{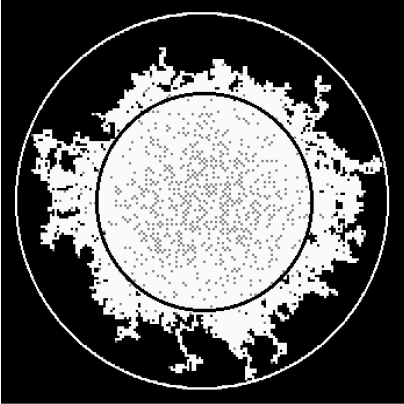}
\end{center}
\caption{Figure from Acedo and Yuste \cite{Yuste}: 
"A snapshot of the set of sites visited by $n = 1000$
random walkers on the two-dimensional lattice. The visited
sites are in white, the unvisited ones are in black and the internal
gray points are the random walkers. The outer white
circle is centered on the starting point of the random walkers
and its radius is the maximum distance from that point
reached by any walker at the time the snapshot was taken.
The internal black circle is concentric with the former but
its radius is the distance between the origin and the nearest
unvisited site."}\label{dendr}
\end{figure}

Let us also make another observation. In the limit of large $n$,
the prefactor for 
the average area of the convex hull of $n$ open Brownian paths, 
namely $\log n$, is identical not only to that 
of the average area of the convex hull of $n$ independent points 
drawn each from a Gaussian distribution $\rho(x,y)\propto 
e^{-\frac{x^2+y^2}{2}}$, but also to that of the number of distinct 
sites visited by $n$ independent random walkers on a 
lattice\footnote{We thank Hern\'an Larralde for drawing our 
attention to this point.}. Indeed the number of distinct sites visited by 
$n$ independent walkers on a lattice (\textit{eg} $\mathbb{Z}^2$) has been 
studied 
systematically by H.~Larralde \textit{et al.} \cite{Larralde, LarrNature} 
(see also \cite{Yuste}). For 
$n$ lattice walks ($n\gg 1$) each of $k$ step 
(steps are only allowed to neighbourings sites), 
Larralde \textit{et al.} have identified three regimes 
according to the value of $k$. For the 
second of these regimes, the intermediate one, the system 
is in a sort of diffusive state so that the number of distinct 
sites visited $\langle S_n(k)\rangle $ grows like the area of the 
disk of radius $\sqrt{k}$, that is, proportionally to $k$, 
with \textbf{a prefactor $\log n$}. In this regime, Acedo and 
Yuste \cite{Yuste} describe the explored territory as "a corona of 
dendritic nature [characterized by filaments created by the 
random walkers wandering in the outer regions] and an inner 
hyperspherical core [where there is much overlapping].\fg\ 
(\textit{cf.}~fig.~\ref{dendr}).
\medskip

The detailed transposition between lattice-walk models and ours ($n$ planar Brownian motions of fixed duration $T$) requires much care,\footnote{In particular because of the transition to 
both continuous time (that is an infinite number of steps) and continuous space (a lattice constant that tends to 0).} but it is interesting to note that in the intermediate regime, the 
number of distinct sites visited by $n$ independent lattice walkers grows like a circle with the same radius as that of the convex hull of $n$ planar Brownian paths.
\bigskip

\section{Conclusion}

The method presented in this paper, based on the use of support functions and 
Cauchy formulae, 
allows one to treat, in a general way, random convex hull problems in a plane, 
whether the 
random points 
considered are independent or correlated. Our work makes an important
link between the two-dimensional convex hull problem and the
subject of extreme value statistics.

We have shown here how this method can be implemented successfully in the case 
of 
independent points in the plane and when the points are correlated
as in the case when the points represent the positions of
a planar Brownian motion of a given fixed duration $T$. This method
should be adaptable to 
treat other types of random paths, such as discrete-time random walks or 
anomalous diffusion processes such as Lévy flights.

In addition, we have shown how to suitably generalise this method
to compute the mean perimeter and the mean area of the global
convex hull of $n$ independent Brownian paths.
Our work leads to several interesting open questions:
\begin{itemize}

\item For example,
can one go beyond the first moment and compute, for instance, the
full distribution of the perimeter and the area of the convex hull
of $n$ independent Brownian paths? 

\item
For a single random walker
of $N$ steps,
the average number of vertices of its convex hull is known
from Baxter's work~\cite{Bax}: $\langle F_N\rangle = 2(1+1/2+1/3+\ldots+1/N)\sim 
2 \log 
(N)$ for large $N$. It remains an outstanding problem to generalise
this result to the case of the convex hull of $n$ independent random walkers 
each of step $N$. 

\item Furthermore, we have only studied the convex hull
of $n$ independent Brownian paths. However, in many situations, the animals
interact with each other leading to collective behavior such as flocking. It 
would thus be very interesting 
to study the effect of interaction between walkers on the statistics of
their convex hull. 

\item Finally, it would be interesting to study the statistics
of the convex 
polytope
associated with Brownian paths (one or more) in $3$ dimensions.
Let us remark that Cauchy's formula
exist in higher dimensions and that it is possible to apply a method
similar to the one developped here in order to compute
the average surface area\footnote{details will be published elsewhere}.
\end{itemize}

\begin{acknowledgements}
We wish to thank D. Dhar , H. Larralde  and B. Teissier for useful discussions.
\end{acknowledgements}

\appendix
\section{Proof of Cauchy's formulae}
\label{CauchyDem}
We give here a quick "proof" of Cauchy's formulae\footnote{We thank Deepak Dhar for suggesting the idea of this demonstration.}:
\begin{equation}
L =\int_{0}^{2\pi} M(\theta)\ d\theta \label{Cau1}
\end{equation}
\begin{equation}
A =\frac{1}{2}\int_{0}^{2\pi} \left( M^2 (\theta) - \left(M'(\theta)\right)^2\right)\ d\theta \label{Cau2}
\end{equation}
\bigskip

We consider a polygonal curve and, without loss of generality, examine the integrals appearing in Cauchy's formulae on a portion of the curve corresponding to the configuration shown on 
figure~\ref{CauchyDem}.
\begin{figure}[h]
\begin{center}
\includegraphics[height=6cm, width=5.85cm]{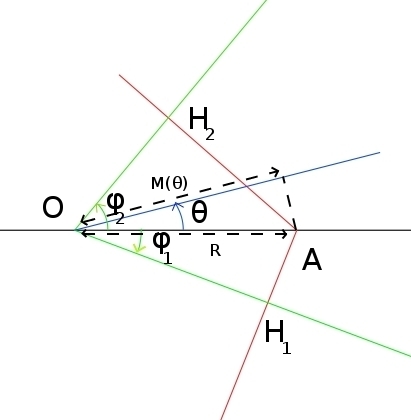}\label{CauchyFig}
\end{center}
\caption{Portion of a polygonal curve (in red) near one of its vertices (A), with the lines through the origin perpendicular to the curve before A and after A (in green), together with the 
line of direction $\theta$ (in blue)}.
\end{figure}
On the interval $\theta \in [-\phi_1,\phi_2]$, the value of the support function of the polygonal curve will be given by $A$. Writing $R$ for the distance between the origin $O$ and vertex 
$A$, we therefore have:
\begin{equation}
M(\theta)=R\cos \theta
\end{equation}
\medskip

The first of Cauchy's formulae then gives:
\begin{equation}
L=\int_{-\phi_1}^{\phi_2}\ M(\theta)\ d\theta = R(\sin \phi_1 + \sin \phi_2),
\end{equation}
which is indeed the length of the curve (that is, $H_1A+AH_2$) between $H_1$ and $H_2$.
\medskip

As for the second of Cauchy's formulae, it gives:
\begin{align*}
A&=\frac{1}{2}\int_{-\phi_1}^{\phi_2}\ \left(\left[M(\theta)\right]^2-\left[M'(\theta)\right]^2\right)\ d\theta \\
&=\frac{R^2}{2}\int_{-\phi_1}^{\phi_2}\ \left[\cos ^2\theta-\sin ^2\theta\right] d\theta\\
&=\frac{R^2}{2}\left(\sin\phi_2\ \cos\phi_2+\sin\phi_1\ \cos\phi_1\right)
\end{align*}
which is indeed the area of the polygon $OH_1AH_2$.
\medskip

This proves the formulae for closed polygonal curves containing the origin. If the origin is outside, one can see that the signs of the various terms will lead to cancellations and the 
formulae will remain valid. Finally, taking the continuous limit yields the result for smooth curves.

\section{Time at which the maximum of $n$ 1-dimensional Brownian motion is attained\label{ToMN}}
Let us write:
\begin{equation}
\text{Prob}(M_n=M,\tau^*=\tau)\equiv \rho_n(\tau,M)\ d\tau\ dM,
\end{equation}
where:
\begin{itemize}

\item $M_n$ is the global maximum of the $n$ Brownian motions,

\item $\tau^*$ is the time at which this global maximum is attained,

\item $\rho_n$ is the joint probability density function of $M_n$ and $\tau^*$.
\end{itemize}
\medskip

This probability can be written as the probability that one of the motions attains its maximum $M$ at time $\tau$ and the $n-1$ others all have a maximum which is less than $M$:
\begin{equation}
\rho_n(\tau,M)= n\rho_1(\tau,M) [F(M)]^{n-1},
\end{equation}
$F(M)$ being, as before, the cumulative distribution function of the maximum of one standard Brownian motion on the interval $[0,T]$:
$$ F(M)= \text{erf}\left( \frac{M}{\sqrt{2T}}\right).$$
\medskip

\noindent The joint probability density function 
$\rho_1(\tau,M)$ of the maximum $M$ and the time $\tau$
at which it happens can be computed 
using various techniques. 
The simplest of them is to use the Feynman-Kac
path integral method, but suitably adapted with a
cut-off~\cite{SMAC,SM}. This technique has recently been used~\cite{RFSM,MKRF} 
to
compute exactly the joint distribution $\rho_1(\tau,M)$
of a single Brownian motion, but subject to
a variety of constraints, such as for a Brownian excursion,
a Brownian meander etc. The results are nontrivial~\cite{MKRF}
and have been recently verified using an alternative functional
renormalization group approach~\cite{GSPL}. For a single free
Brownian motion (the case here), this method can be similarly
used and it provides a simple and compact result 
$$\rho_1(\tau,M)=\frac{M}{\pi\tau^\frac{3}{2}\sqrt{T-\tau}}
e^{-\frac{M^2}{2\tau}}.$$
\medskip

\noindent We then obtain the marginal  
distribution $\rho_n(\tau)$ by integrating out $M$. It has
the scaling form $\rho_n(\tau)=\frac{1}{T}f_n\left(\frac{\tau}{T}\right)$
where the scaling function $f_n(z)$ is given in Eq. (\ref{maxtN}) and is
plotted, for various values of $n$, in Fig. (\ref{DTMB}).

For Brownian bridges, the reasoning is exactly the same, but of course $F(M)$ and $\rho_1$ differ:
\begin{eqnarray}
\rho_1 (\tau, M) &=& \frac{2T}{\pi}\frac{M^2}{[\tau(T-\tau]\frac{3}{2}}\ e^{-\frac{M^2T}{2\tau(T-\tau)}}\\
F(M)&=&1-e^{-\frac{2M^2}{T}},
\end{eqnarray}
which can both be derived using the technique we pointed to above \cite{ACSM1, SM, RFSM, MKRF}.
\medskip

\noindent We then obtain $\rho_n(\tau)= 
\frac{1}{T}g_n\left(\frac{\tau}{T}\right)$ where
\begin{equation}
g_n(z) = \frac{4n}{\sqrt{\pi}}\int_0^\infty\ u^2\ e^{-u^2}\ 
\left[1-e^{-4u^2\,z\,(1-z)}\right]^{n-1}\ du.
\end{equation}
Figure~\ref{DTPB} shows 
a plot of $g_n(z)$ for different values of $n$.

\section{Asymptotic behaviour\label{AsBe}}
\subsection{Open paths - average perimeter}\label{AsLNMB}
Letting $M_n$ be the maximum of $n$ independent Brownian paths each of duration $T$, 
we 
write $\mathrm{Prob}(M_n\leq M) \equiv  F_n(M)$. This can be expressed 
in terms of the cumulative distribution function $F_1(M)$ of 
the maximum of a standard one-dimensional Brownian motion:
\begin{eqnarray}
F_n(M) &=& [F(M)]^n \nonumber\\
&=&\left[ \frac{2}{\sqrt{\pi}}\int_0^\frac{M}{\sqrt{2T}}\ e^{-u^2}\ du\right]^n\nonumber\\
&=& \left[ 1-\frac{2}{\sqrt{\pi}}\int_\frac{M}{\sqrt{2T}}^\infty\ e^{-u^2}\ du\right]^n\label{E1}
\end{eqnarray}
In the limit when $n$ and $M$ become very large, (\ref{E1}) becomes:
\begin{eqnarray}
F_n(M)&\sim & \exp \left[ n\ln \left( 1-\frac{2}{\sqrt{\pi}}\int_\frac{M}{\sqrt{2T}}^\infty\ e^{-u^2}\ du\right)\right]\nonumber\\
&\sim & \exp \left[-n\frac{2}{\sqrt{\pi}}\int_\frac{M}{\sqrt{2T}}^\infty\ e^{-u^2}\ du\right]\label{E2}
\end{eqnarray}
Integrating by parts yields:
\begin{equation}
\int_\frac{M}{\sqrt{2T}}^\infty\ e^{-u^2}\ du = 
\frac{\sqrt{T}\,e^{-\frac{M^2}{2T}}}{\sqrt{2}M} +O(1)
\end{equation}
Inserting this into equation (\ref{E2}), one obtains:
\begin{eqnarray}
F_n(M) &\sim & e^{\frac{-2\,n\,\sqrt{T}\,e^{-M^2/2T}}{M\sqrt{2\pi}}}\\
&\sim & e^{-e^{-\frac{1}{2T}\left(M^2-2T\ln n\right)}}\label{E3}
\end{eqnarray}
Here we write: $$\delta\equiv M-\sqrt{2T\ln n},$$ assuming that $\delta$ vanishes at large $n$, as we will be able to check \textit{a posteriori}. It follows that:
$$ M^2 = 2T \ln n \left( 1+\frac{\delta}{\sqrt{2T\ln n}}\right)^2 \sim 2T\ln n \left( 1+\frac{2\delta}{\sqrt{2T\ln n}}\right)$$
Hence:
$$M^2 - 2T \ln n \sim 2\delta \sqrt{2T\ln n} = 2\sqrt{2T\ln n} \left( M-\sqrt{2T\ln n}\right)$$
Inserting this into (\ref{E3}) yields:
\begin{equation}
F_n(M) \sim e^{-e^{-\sqrt{\frac{2\ln n}{T}}\left( M-\sqrt{2T\ln n}\right)}}\label{E4}
\end{equation}

Now, to compute the asymptotic behaviour of $M_n$ when $n$ is large, let us start from:
\begin{eqnarray}
\langle M_n \rangle & =& \int_0^\infty M F_n'(M)\ dM \nonumber\\
&\sim & \int_A^\infty MF_n'(M)\ dM
\end{eqnarray}
where $A>>1$ will "disappear" at a later stage.\\
Combining this with Eq.~(\ref{E4}), one obtains:
\begin{equation}
\langle M_n \rangle \sim \int_A^\infty\ M\ \sqrt{\frac{2\ln n}{T}}e^{-\sqrt{\frac{2\ln n}{T}}\left( M-\sqrt{2T\ln n}\right)}e^{-e^{-\sqrt{\frac{2\log n}{T}}\left( M-\sqrt{2T\ln n}\right)}}\ dM\label{E5}
\end{equation}
Setting: $$y=\sqrt{\frac{2\log n}{T}}\left( M-\sqrt{2T\ln n}\right)$$ 
one arrives at:
\begin{equation}
\langle M_n \rangle \sim 
\int_{\sqrt{\frac{2\ln n}{T}}
\left( A-\sqrt{2T\ln n}\right)}^\infty\ 
\left( \sqrt{2T\ln n} +y\sqrt{\frac{T}{2\ln n}}\right)\,e^{-y}\, e^{-e^{-y}}\ dy
\end{equation}
In the limit when $n\rightarrow \infty$, this leads to:
\begin{eqnarray}
\langle M_n \rangle &\sim &\int_{-\infty}^\infty\ \sqrt{2T\ln n}\,e^{-y}\, 
e^{-e^{-y}}\ dy\nonumber\\
 &\sim & \sqrt{2T\ln n}
\end{eqnarray}
Thus, the average perimeter of the convex hull of $n$ Brownian paths in the 
plane behaves for large $n$ 
\begin{eqnarray}
\langle L_n \rangle &\sim & 2\pi \sqrt{2T\ln n}.
\end{eqnarray}

\subsection{Open paths: average area}\label{AsANMB}
We wish to compute the asymptotical behaviour (for $n$ large) of the average area of the convex hull of $n$ open Brownian paths in the plane, all independent and of duration $T$. We apply Cauchy's formula (\ref{IAC2}) in the context of isotropic samples:
\begin{equation}
\langle A_n\rangle = \pi \left( \langle M_n^2 \rangle - \langle
[M'_n]^2\rangle\right). 
\end{equation}
The first term of the right-hand side is easily computed (it suffices to substitute 
$M^2$ for $M$ in Eq.~(\ref{E5})):
\begin{equation}
\langle M_n^2 \rangle \sim 2T\ln n
\end{equation}

To obtain (\ref{bngd}), it thus remains to show that $\langle A_n\rangle$ is 
dominated by $\langle M_n^2 \rangle$. Recall Eq.~(\ref{R3}):
\begin{equation}
\langle A_n\rangle= {4nT}\,{\sqrt{\pi }}\, \int _0 ^\infty du\ u\
\left[\text{erf}(u)\right]^{n-1}\left(u e^{-u^2}-g(u)\right),\label{RA1}
\end{equation}
where $$g(u)=\frac{1}{2\sqrt{\pi}}\int_0^1 \frac{e^{-{u^2}/{\tau}}\
dt}{\sqrt{\tau(1-\tau)}}.$$
The integral in Eq.~(\ref{RA1}) is dominated by the contribution from the large-$u$ 
part. Therefore, we examine in this limit:
\begin{equation}
\int_0^1 d\tau \frac{1}{\sqrt{\tau(1-\tau)}}e^{-\frac{u^2}{\tau}}
\end{equation}
Setting $\tau=1-y$:
\begin{eqnarray}
\int_0^1 d\tau \frac{1}{\sqrt{\tau (1-\tau)}}e^{-\frac{u^2}{\tau}}&=&\int_0^1\ dy \frac{1}{\sqrt{y(1-y)}}e^{-\frac{u^2}{1-y}}\\
&\sim & \int_0^1\ dy \frac{1}{\sqrt{y(1-y)}}e^{-u^2(1+y)}\\
&\sim & e^{-u^2}\int_0^1\ dy \frac{1}{\sqrt{y(1-y)}}e^{-yu^2}
\end{eqnarray}
We now write: $$z=u\sqrt{y}.$$
This leads to:
\begin{eqnarray*}
\int_0^1 d\tau \frac{1}{\sqrt{\tau (1-\tau)}}e^{-\frac{u^2}{\tau}}&\sim & e^{-u^2}\int_0^ u\ \frac{2z\ dz}{u^2} \frac{1}{\sqrt{\frac{z^2}{u^2}(1-\frac{z^2}{u^2})}}e^{-z^2}\\
&\sim & \frac{2e^{-u^2}}{u}\int _0^u\ dz \frac{1}{\sqrt{1-\frac{z^2}{u^2}}}e^{-z^2}\\
&\sim & \frac{2e^{-u^2}}{u}\int _0^u\ dz (1+\frac{z^2}{2u^2})e^{-z^2}\\
&\sim & \frac{2e^{-u^2}}{u}\sqrt{\frac{\pi}{2}} \text{erf}(u)+\frac{e^{-u^2}}{u}\int _0^x\ dz\ \frac{z^2e^{-z^2}}{u^2}
\end{eqnarray*}
An integration by parts shows that the second term on the right-hand side is  $o\left( \frac{e^{-u^2}}{u^2}\right)$, whence:
\begin{equation}
\int_0^1 d\tau \frac{1}{\sqrt{\tau(1-\tau)}}e^{-\frac{u^2}{\tau}}\sim \frac{\sqrt{\pi}e^{-u^2}}{u}.
\end{equation}
Returning to Eq. (\ref{RA1}), where the integral is dominated by the 
contribution 
from the large $u$ part, we have:
\begin{eqnarray}
\langle A_n\rangle &=& {4nT}\,{\sqrt{\pi }}\, \int _0 ^\infty du\ u\
\left[\text{erf}(u)\right]^{n-1}\left(u e^{-u^2}-g(u)\right)\\
&\sim& {4nT}\,{\sqrt{\pi }}\, \int _0 ^\infty du\ u\
\left[\text{erf}(u)\right]^{n-1}\left(u 
e^{-u^2}-\frac{e^{-u^2}}{2u}\right)\\
&\sim& {4nT}\,{\sqrt{\pi }}\, \int _0 ^\infty du\ u^2\
\left[\text{erf}(u)\right]^{n-1} e^{-u^2}\\
&\sim & \pi \langle M_n^2 \rangle \\
&\sim& 2\pi T\ln n.
\end{eqnarray}

\subsection{Closed paths: average perimeter}\label{AsLNPB}
The calculation is analogous to that of subsection \ref{AsLNMB}; indeed, the only 
difference is that the cumulative distribution function $F_n$ of the maximum $M$ of $n$ one-dimensional Brownian bridges is given by:
\begin{equation}
F_n(M)=[1-e^{-\frac{2M^2}{T}}]^n
\end{equation}
In the limit when $n$ and $M$ become very large, we then have:
\begin{eqnarray}
F_n(M)&\sim & \exp \left[ n\ln \left( 1-e^{-\frac{2M^2}{T}}\right)\right]\nonumber\\
&\sim & e^{-e^{-\frac{2}{T}\left(M^2-\frac{T}{2}\ln n\right)}}
\end{eqnarray}
We retrieve here the same equation as~(\ref{E3}), where $T$ is replaced by~$\frac{T}{4}$. We can thus deduce the result given in~(\ref{ancgd}): 
\begin{eqnarray}
\langle L_n \rangle &\sim & \pi \sqrt{2T\ln n}.
\end{eqnarray}

\subsection{Closed paths: average area}\label{AsANPB}
We wish to compute the asymptotic behaviour (for $n$ large) 
of the average area of the convex hull of $n$ closed 
Brownian paths in the plane, all independent and of duration $T$. 
The analysis is similar to that of subsection \ref{AsANMB}. We apply Cauchy's 
formula (\ref{IAC2}) in the context of isotropic samples:
\begin{equation}
\langle A_n\rangle = \pi \left( \langle M_n^2 \rangle - \langle
[M'_n]^2\rangle\right). 
\end{equation}
The first term on the right-hand side is, as before, easily computed from 
subsections~\ref{AsLNPB} and \ref{AsLNMB}:
\begin{equation}
\langle M_n^2 \rangle \sim \frac{T}{2}\ln n
\end{equation}

We now show that $\langle A_n\rangle$ is dominated by $\langle M_n^2 \rangle$. The 
equivalent of Eq.~(\ref{RA1}) for closed paths is:
\begin{equation}
\langle A_n\rangle= \frac{2nT}{\sqrt{\pi}}\ \int _0^\infty \ du\ u^2\left(1-e^{-u^2}\right)^{n-1}\left(ue^{-u^2}-g(u)\right),\label{RA2}
\end{equation}
where: $$g(u)=\frac{1}{8}\int_0^1\ d\tau\ \frac{e^{-\frac{u^2}{4\tau(1-\tau)}}}{\sqrt{\tau(1-\tau)}}.$$
The integral in Eq.~(\ref{RA2}) is dominated by the contribution from the large-$u$ 
part. Therefore, we examine $g(u)$ in this limit.\\
\begin{eqnarray}
g(u)&=&\frac{1}{8}\int_0^1\ d\tau\ \frac{e^{-\frac{u^2}{4\tau(1-\tau)}}}{\sqrt{\tau(1-\tau)}}\\
&=&\frac{1}{4}\int_0^{\frac{1}{2}}\ d\tau\ \frac{e^{-\frac{u^2}{4\tau(1-\tau)}}}{\sqrt{\tau(1-\tau)}}
\end{eqnarray}
We now write: $$\tau=\frac{1}{2}-z.$$
This leads to:
\begin{eqnarray*}
\frac{1}{4}\int_0^{\frac{1}{2}}\ d\tau\ \frac{e^{-\frac{u^2}{4\tau(1-\tau)}}}{\sqrt{\tau(1-\tau)}}&= &\frac{1}{2}\int_0^{\frac{1}{2}}\ dz\ \frac{e^{-\frac{u^2}{1-4z^2}}}{\sqrt{1-4z^2}}
\end{eqnarray*}
Setting: $$v=2uz,$$ then yields:
\begin{eqnarray*}
\frac{1}{2}\int_0^{\frac{1}{2}}\ dz\ 
\frac{e^{-\frac{u^2}{1-4z^2}}}{\sqrt{1-4z^2}} 
&= & \frac{1}{4}\int_0^{u}\ \frac{dv}{u}\ 
\frac{e^{-\frac{u^2}{1-\frac{v^2}{u^2}}}}{\sqrt{1-\frac{v^2}{u^2}}}\\
&\sim & \frac{1}{4}\frac{e^{-u^2}}{u}\int_0^{u}\ dv\ 
e^{-v^2}(1+\frac{v^2}{2u^2})\\
&\sim & \frac{e^{-u^2}}{8u}\sqrt{\pi}\ 
\text{erf}(u)+O\left(\frac{e^{-u^2}}{u^2}\right)
\end{eqnarray*}

Returning to Eq. (\ref{RA2}), where the integral is dominated by the contribution 
from the large-$u$ part, we therefore have:
\begin{eqnarray}
\langle A_n\rangle &=& \frac{2nT}{\sqrt{\pi}}\ \int _0^\infty \ du\ u^2\left(1-e^{-u^2}\right)^{n-1}\left(ue^{-u^2}-g(u)\right)\\
&\sim& \frac{2nT}{\sqrt{\pi}}\ \int _0^\infty \ du\ u^2\left(1-e^{-u^2}\right)^{n-1}\left(ue^{-u^2}-\frac{\sqrt{\pi}e^{-u^2}}{8u}\right)\\
&\sim& \frac{2nT}{\sqrt{\pi}}\ \int _0^\infty \ du\ u^3\left(1-e^{-u^2}\right)^{n-1}e^{-u^2}\\
&\sim & \pi \langle M_n^2 \rangle\\
& \sim & \frac{\pi T}{2}\ln n.
\end{eqnarray}

\bibliographystyle{spmpsci}
\bibliography{longhull6}

\end{document}